\documentclass[12pt]{article}

\usepackage{siunitx} 
\usepackage[margin=1in]{geometry}
\usepackage{amsmath}
\numberwithin{equation}{section}
\usepackage{graphicx}
\usepackage{natbib}
\usepackage{underscore}
\usepackage{physics}
\usepackage[ruled,vlined]{algorithm2e}
\usepackage{algpseudocode}
\usepackage{booktabs}
\usepackage{amsfonts}
\usepackage{bm}
\usepackage{threeparttable}
\usepackage{rotating}
\usepackage{setspace}
\usepackage{subcaption}
\usepackage[width=.95\textwidth, skip=0.1\baselineskip]{caption}
\usepackage[colorlinks, citecolor=blue, urlcolor=blue, linkcolor=blue]{hyperref}
\usepackage{xcolor}
\usepackage[]{lineno}
\allowdisplaybreaks

\allowdisplaybreaks 

\let\proglang=\textsf
\newcommand{\pkg}[1]{{\texttt #1}}

\usepackage{multirow, makecell}
\usepackage{booktabs,array}
\usepackage[affil-it]{authblk}

\title{Unobserved Heterogeneity in Threshold Regression Based on the
  Hitting Times of a Reflected Brownian Motion for Recurrent
  Hypoglycemia}

\date{}

\author[1,3]{Yingfa Xie}
\author[2]{Haoda Fu}
\author[3]{Yuan Huang  \thanks{Email address: \texttt{yuan.huang@yale.edu}; 
	corresponding author}}
\author[1]{Jun Yan}
\affil[1]{Department of Statistics, University of Connecticut}
\affil[2]{Indiana University School of Medicine}
\affil[3]{Department of Biostatistics, Yale School of Public Health}

\begin{document}
\maketitle

\begin{abstract}
Analyses of recurrent hypoglycemia are critical for effective
treatment management in diabetic patients. Typically, within-subject
dependency in such analyses is captured through subject-level frailty.
Recent research has modeled recurrent hypoglycemia using the first
hitting times of a reflected Brownian motion. A close examination of
this approach reveals that it does not adequately account for varying
frailties  among individuals, which indicate notable heterogeneity. To
address this gap, we propose a finite mixture model of the first
hitting time distribution of the reflected Brownian motion. This model
allows for component-specific regression coefficients and frailty
parameters, providing nuanced insights into how risk factors
differently affect patient subgroups. We employ a Bayesian framework
for inference, utilizing Markov chain Monte Carlo for estimation.
Model selection is conducted using the Deviance Information Criterion
and the Logarithm of the Pseudo-Marginal Likelihood. The effectiveness
of these criteria is assessed through simulation studies. 
Application to recurrent hypoglycemia modeling revealed two subgroups
with different risk profiles, as reflected in their volatilities.
Bayesian model comparison criteria favor the model with component
specific regression coefficients for volatilities. The subgroup with
lower volatility exhibits a larger variance and, hence, a greater
level of heterogeneity.

\bigskip


\end{abstract}

\doublespacing

\newpage

\section{Introduction}

Hypoglycemia represents a significant adverse effect in glucose
management and is a focal point of diabetes clinical research
\citep{fu2016hypoglycemic, ma2021heterogeneous}. Patients with
diabetes display diverse reactions or responses to their treatment
regimens. For instance, in the 24-week DURABLE trial
\citep{buse2009durability}, experiences of hypoglycemia among 
patients varied widely: some faced more than 100 episodes, 
underscoring it as a potential adverse effect of glucose management, 
while others seldom experienced any. 
This marked variability in hypoglycemia risk highlights the presence
of substantial unobserved heterogeneity within the patient population.
Tailoring treatment plans to specific patient subgroups, rather than
adopting a ``one size fits all'' approach, may offer more effective
diabetes management strategies \citep{qu2022identifying,
jiang2018clinical}.

In practical healthcare settings, hypoglycemia episodes are typically
self-reported by patients,  facilitated by the occurrence of distinct 
symptoms such as headaches, hunger, and rapid heartbeat that serve as 
clear indicators of critically low glucose levels
\citep{cryer2003hypoglycemia}. The pronounced and prompt manifestation
of these symptoms enables patients to identify and report hypoglycemia
episodes, thereby supporting accurate and timely documentation of such
events. On the other hand, hyperglycemia,
characterized by elevated glucose levels, often lacks overt symptoms
that are immediately recognizable to the patient, making its detection
more complex and generally requiring clinical intervention through
blood tests. Given these significant differences in the observability
of symptoms between hypoglycemia and hyperglycemia, the DURABLE trial
\citep{buse2009durability} specifically opted to gather data only on
hypoglycemic events that could be reliably reported by patients.

The First Hitting Time (FHT) model
\citep{whitmore1986first,lee2003first, aalen2001understanding} offers
a refined approach for modeling time-to-event data, particularly in
medical scenarios where it is crucial to predict the timing of specific
health events. This
model conceptualizes events as the moment an underlying stochastic
process, such as a patient's evolving health condition, crosses a
predefined threshold \citep{lee2019survey}, which could signify a
critical health event like hypoglycemia. 
Threshold regression, which incorporates covariate effects into model
parameters such as the volatility of a Wiener process, enhances the
precision of the FHT model. By accounting for individual patient
characteristics, it provides more accurate predictions of event times
\citep{lee2006threshold}. 
The challenge of modeling unobserved heterogeneity is commonly
addressed by incorporating random effects
\citep{pennell2010bayesian, malefaki2015modelling, economou2015bayesian}
or mixture components
\citep{whitmore2007modeling, lee2008threshold, race2021semi}
through their respective modifications of the FHT model.
As this work focuses on mixture threshold regression models, a
comprehensive review of recurrent event modeling is beyond the scope
of this paper; we refer interested readers to recent reviews for the
topics \citep{cook2007statistical, charles2019analyze}.

Applying the FHT model to hypoglycemia involves specific challenges,
particularly in accurately modeling the dynamics of blood glucose
levels. Blood glucose fluctuations can naturally be represented by
Brownian motion, reflecting the unpredictable changes influenced by
insulin administration, dietary intake, and physical activity. This
stochastic representation is key to understanding the risk and
occurrence of hypoglycemic events, marked by glucose levels crossing a
lower boundary indicative of critically low glucose levels. Earlier
studies have adapted the FHT model for recurrent hypoglycemic events,
employing sequences of Brownian motions with reflective barriers to
account for physiological expectations that glucose levels will
eventually decrease, often due to the action of glucose-lowering
medications \citep{xie2025recurrent}.
However, the detection of a bi-modal
distribution in the fitted subject-level random effects suggests the
existence of significant heterogeneity in the patient population,
underscoring the need for a more nuanced modeling approach.

To address this gap, we propose a finite
mixture FHT model tailored to the reflected Brownian
motion process to capture the unobserved heterogeneity among
patients with recurrent hypoglycemic events. By incorporating
subject-level covariates and random effects, the model differentiates
patient risk profiles, enabling a more individualized analysis of
hypoglycemia risk. This approach offers insights into the impacts of
different heterogeneity on recurrent hypoglycemic events, marking a
significant contribution to personalized diabetes
care. Model parameters are estimated in a Bayesian framework with
Markov chain Monte Carlo (MCMC). Different distributional
specifications of the random effects are evaluated via model selection
criteria. Through extensive simulation studies and application to a
motivating data, our research broadens the application of the FHT
model, providing novel opportunities for a more comprehensive 
understanding and improved management of diabetes.
The proposed methodology extends the non-mixture recurrent event model
based on the FHT of reflected Brownian motion developed in
\citet{xie2025recurrent} to a finite mixture model for capturing
additional unobserved heterogeneity in recurrent hypoglycemic events.

The structure of the remainder of this paper is as follows.
Section~\ref{sec:data} introduces the analysis data based on the 
DURABLE study and discuss existing studies. In
Section~\ref{sec:model}, a mixture FHT model of reflected Brownian
motion is proposed and Bayesian parameter estimation and model
selection methods are presented. An application of the methods is
given in Section~\ref{sec:aplc}. Simulation studies that validate the
Bayesian inferences are reported in Section~\ref{sec:simulation}. A
discussion is provided in Section~\ref{sec:disc}.

\section{Data and Challenge} \label{sec:data}

\begin{figure} [tbp]
\centering
\includegraphics[width=\linewidth]{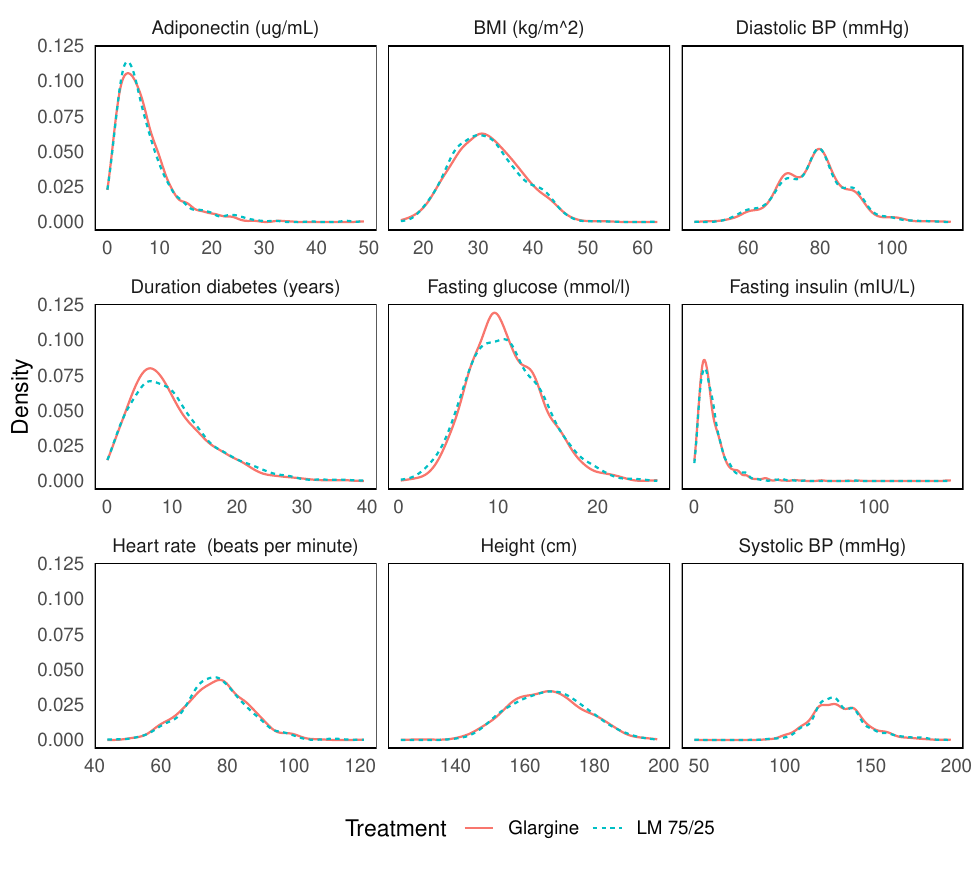}
\caption{Distributions for continuous covariates by insulin 
starter regimens.}
\label{fig:datDescript}
\end{figure}

The DURABLE trial \citep{buse2009durability} compared the efficacy of
two starter insulin regimens, insulin lispro mix 75/25 versus insulin
glargine, in achieving glycemic control for type 2 diabetes patients.
The study enrolled $2187$ patients from $11$ countries. During the
trial, the timing (in days) of hypoglycemic events was recorded for
each patient over their follow-up period, which had a median duration
of 168 days. 
We analyzed a synthetic dataset derived from the original DURABLE
trial, in which covariates were lightly perturbed while the original
event gap times were preserved. This ensures privacy protection while
retaining the key features that motivate the current study.
The dataset includes baseline covariates for each patient,
such as fasting insulin, adiponectin, fasting blood
glucose, height, body mass index (BMI), heart rate, diastolic and
systolic blood pressure,  and duration of diabetes, which are crucial
for identifying risk factors associated with hypoglycemic events.
Figure~\ref{fig:datDescript} displays the distributions of continuous
baseline covariates by treatment group, showing little difference
between the two regimens at baseline. 
Descriptive statistics of the 
continuous covariates are presented in Section~1 of Supplementary
Materials.
The dataset also includes two categorical
variables: one indicating the type of oral antihyperglycemic drugs
used---categorized as thiazolidinedione, sulfonylurea, or both---and
another denoting the assigned starter insulin regimen, comparing
twice-daily lispro mix 75/25 (LM75/25; 75\% lispro protamine
suspension and 25\% lispro) with once-daily insulin glargine.

The dataset for our study contained $n = 1943$
patients after excluding the subjects with missing value or data
points outside the reference range. To prepare for the analysis, we
log-transformed adiponectin level and fasting insulin level, which
have extremely high values, and standardized all the continuous
covariates. Among these patients, $1207$ ($62$\%) received
sulfonylurea alone, $166$ ($9$\%) received thiazolidinedione alone,
and $570$ ($29$\%) received both oral antihyperglycemic drugs.
For ease of
discussion, we set the group receiving both drugs as the reference
group. Two dummy variables, ``sulf-Only'' (equal to 1 if only
sulfonylurea was received) and ``tzd-only'' (equal to 1 if only
thiazolidinedione was received), were created. An indicator variable,
denoted as ``LM'', is used to signify the type of insulin starter
regimen, with a value of 1 assigned to the $959$ $(49\%)$ patients 
receiving LM 75/25 and 0 to the $984$ $(51\%)$ patients administered
glargine.

The daily hypoglycemic event rates exhibited substantial variability,
ranging from $0$ to $0.77$, with an average rate of $0.07$. There
were $608$ patients experienced more than one hypoglycemic event
within a single day. This variability highlights the potential need
for our proposed model to account for the diverse risk profiles within
the diabetic patient cohort.

We briefly review the modeling framework proposed in
\citet{xie2025recurrent}, which serves as the foundation for the
mixture extension developed in this paper. 
\citet{xie2025recurrent} modeled hypoglycemic events by the FHT of 
a reflected Brownian motion. This approach conceptualizes the
recurrent hypoglycemic events as a sequence of Brownian motion
reflected from above, each hitting a lower boundary, followed by a reset 
of the process at a predetermined level. The upper reflection barrier
of the Brownian motion allows bypassing the need for capturing
hyperglycemic event times, which are often unreliable or unobservable
in self-reported datasets. The gap times between the successive events
of the same patient are assumed to be independent conditional on a
subject-specific frailty, which follow a FHT distribution of the
associated reflected Brownian motion. The distribution and density 
functions investigated by \citet{hu2012hitting} are presented in 
Appendix~\ref{sec:FHTdist} for completeness.
Random number generation from the
distribution can be developed with a rejection sampling algorithm. 
In the modeling framework, the subject-level frailty and covariates
were linked to the volatility and the upper reflecting barrier of the
Brownian motion.

\begin{figure} [tbp]
\centering
\includegraphics[width=\linewidth]{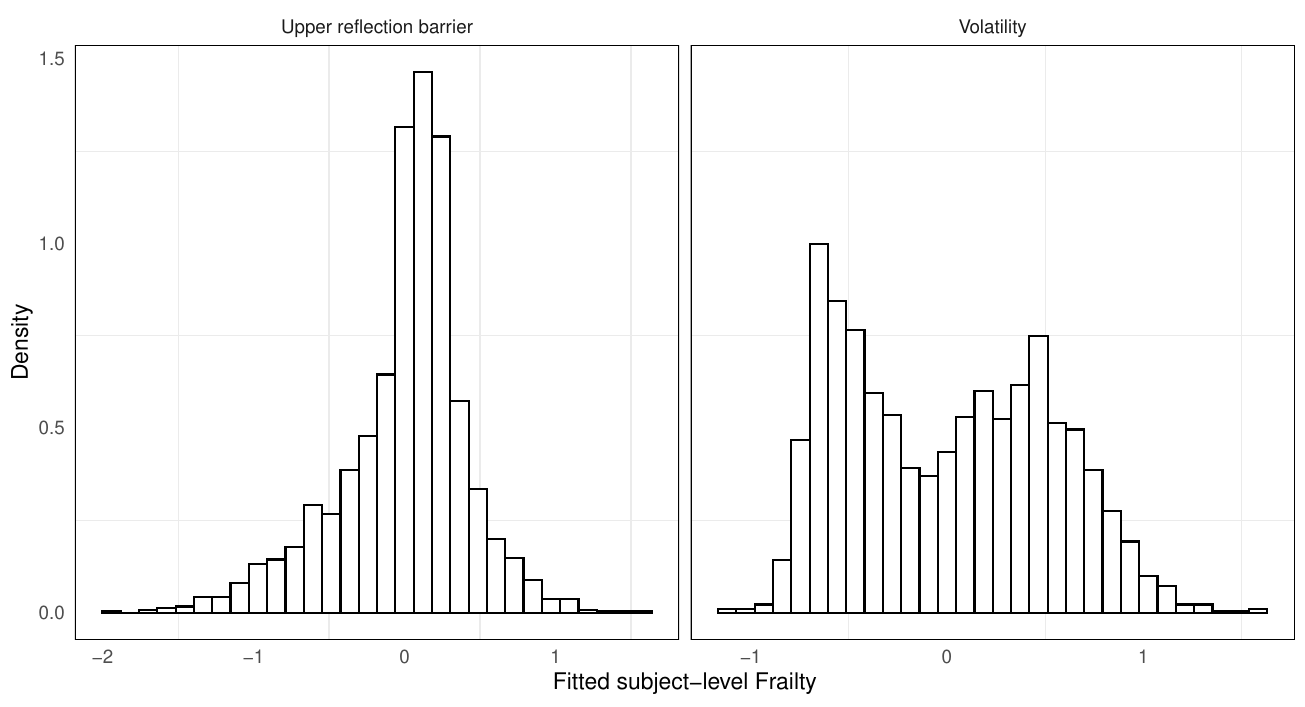}
\caption{The histogram of the posterior mean of the fitted subject-level
	frailty in upper reflecting barrier (left) and volatility (right)}
\label{fig:frailties}
\end{figure}

Incorporating subject-level frailty into the model is crucial for
capturing heterogeneity among patients. The estimated frailty variance
in the volatility model was $0.535$ with 95\% credible interval
$(0.450, 0.620)$, which significantly deviates from zero
\citep{xie2025recurrent}. This result reflects the heterogeneity in
hypoglycemic event rates among patients. Further scrutiny of the
fitted subject-level frailty reveals a bi-modal pattern in the
histogram as
shown in Figure~\ref{fig:frailties}. This pattern suggests unobserved
heterogeneity beyond covariates and subject-level frailty, which
might be captured by a mixture model. Mixture models are commonly used
to handle unobserved heterogeneity among patients from different
subpopulations. Covariate coefficients can be constructed as
component-specific, which allows for diverse effects of the treatments
on patients from different groups. Furthermore, positing that
frailties for patients from different subgroups adhere to different
distributions emerges as a reasonable and logical step forward. Given
these considerations, we advance the framework of
\citet{xie2025recurrent} to a finite mixture FHT model tailored to the
analysis of recurrent hypoglycemic events, thereby aiming to more
accurately capture the intricate heterogeneity within the patient
data.

\section{Methods} \label{sec:model} 

This section introduces the finite mixture FHT model, presents its
Bayesian formulation, and outlines the associated reduced models and
model selection criteria.

\subsection{A FHT Mixture Model for Recurrent Events} \label{subsec:modformula}

We propose a mixture FHT frailty model with $K$ components to further
capture the heterogeneity among subgroups. To address the specific
motivating problem presented in Section~\ref{sec:data}, we consider
the volatility $\sigma$ of the Brownian motion to be
component-specific, with $K = 2$. 
Suppose that subject~$i$ belongs to the $k$th component of the mixture
model with weight~$\rho_k$, where $\sum_{k=1}^2 \rho_k = 1$.
Denoting $\bm{\sigma}_i = (\sigma_{i1}, \sigma_{i2})$,
$\bm{\rho} = (\rho_{1}, \rho_{2})$, the $j$th~gap time~$t_{ij}$ of subject~$i$
is assumed to be drawn from the mixture density
\begin{equation*} 
  p(t_{ij}| \bm{\rho}, x_0, \nu, \kappa_i, \bm{\sigma_i} ) =  
  \sum_{k = 1} ^{2} \rho_{k} f(t_{ij} | x_0, \nu, \kappa_{i}, \sigma_{ik}),
\end{equation*}
where $f(\cdot)$ denotes the FHT density function defined in
\eqref{eq:dens}, and $x_0$ and $\nu$ represent the preset starting
point and the lower boundary of the Brownian motion, respectively,
$\kappa_{i}$ is the upper reflection barrier of the Brownian motion
for subject~$i$, and $\sigma_{ik}$ represents the volatility of the
Brownian motion for subject~$i$ in the $k$th component. Typically, a
smaller upper reflection barrier or a larger volatility is associated
with an increased risk of hypoglycemic events.

Subject-level covariate and frailty are incorporated into the upper reflection
barrier~$\kappa_{i}$ and the volatility~$\sigma_{ik}$ with logarithmic
link. Let $\bm{X}_i = (1, X_{i1}, \cdots, X_{ip})^{\top}$ be a
time-independent covariate vector of subject~$i$.
The models for $\sigma_{ik}$ and $\kappa_{i}$ are given by
\begin{equation} \label{eq:mixmodelpara}
  \log(\sigma_{ik}) = \boldsymbol{X}_i^{\top} \bm{\beta_k} + Z_{1ik}, 
  \qquad \mbox{and} \qquad
  \log(\kappa_{i} - x_0) = \boldsymbol{X}_i^{\top} \bm{\alpha} + Z_{2i}, 
\end{equation}
where $\bm{\alpha} = (\alpha_0, \alpha_1, \cdots, \alpha_{p})^{\top}$
is the regression coefficient vector in upper reflection barrier;
$\bm{\beta_k} = (\beta_{k0}, \beta_{k1}, \cdots, \beta_{kp})^{\top}$,
$k = 1, 2$,
is the component-specific regression coefficient vector for the
volatility. $Z_{1ik}$ and $Z_{2i}$ are the independent random effects
following normal distribution with mean $0$ and variance $\theta_{1k}$
and $\theta_{2}$, respectively. Only the volatility is allowed to have
component-specific regression coefficients, as extending this
flexibility to the upper reflection barrier would substantially
increase model complexity and hinder MCMC convergence.

\subsection{Likelihood, prior, and posterior} \label{subsec:lklh}
Assume that hypoglycemic event times are observed from $n$ patients
during their follow-up periods. For subject~$i$, $i = 1, \cdots, n$,
let $T_{ij}$, $j = 1, \cdots, n_i$, denote the observed time of the
$j$th observation, where the final observation is censored.
The gap times between two successive events are denoted as $t_{ij} =
T_{ij} - T_{i(j-1)}$, where $T_{i0} = 0$. For notation generality,
define the event indicator $\delta_{ij} = 1$ for $j \ne n_i$,
indicating an event observation, and~$0$ otherwise indicating a
censored observation.
That is, the only censored observations are the final one for each
subject. For subject~$i$, the observed data is given by 
$\bm{D}_{i} = \{\bm{X}_i, \bm{t}_i, \bm{\delta}_i\}$, 
where $\bm{t}_i = (t_{i1}, \cdots, t_{in_i})^{\top}$ and
$\bm{\delta}_i = (\delta_{i1}, \cdots, \delta_{in_i})^{\top}$. 
The observed data for all the subjects are denoted by
$\bm{D} = (\bm{D}_1, \cdots, \bm{D}_n)$.

Given the covariate vector~$\bm{X}_i$, the latent membership~$m_i$,
and subject-level frailties~$\bm{z}_i = (z_{1im_i}, z_{2i})$, the gap
times of subject~$i$ are assumed to be conditionally independent. 
Let $\bm{\Omega} = (\bm{\alpha}^{\top}, \bm{\beta}_1^{\top}, \bm{\beta}_2^{\top})^{\top}$, 
the conditional likelihood contribution of subject~$i$ given the
frailties and the membership can be written as
\begin{equation*}
L_{i} (\bm{\Omega} | \bm{D}_i, \bm{z}_i, m_i)
    = \prod_{j = 1}^{n_i}
	\left[f(t_{ij}|x_0, \nu, \kappa_{i}, \sigma_{im_i}) 
		\right]^{\delta_{ij}}
	\left[1 - F(t_{ij}|x_0, \nu, \kappa_{i}, \sigma_{im_i}) 
		\right]^{ (1 - \delta_{ij})},
\end{equation*}
where the density $f$ and distribution $F$ functions are given 
in~\eqref{eq:dens} 
and \eqref{eq:dist}, respectively, and $\sigma_{im_i}$ and $\kappa_{i}$ are defined
in~\eqref{eq:mixmodelpara}.

The recorded event times are usually discrete in the unit of days. In
this case, the gap time $t_{ij}$'s are handled as interval-censored.
The likelihood contribution for subject~$i$ can then be rewritten as
\begin{align}  \label{eq:intvcondlklksubj}
L_{i} (\bm{\Omega} | \bm{D}_i, \bm{z}_i, m_i)
= \prod_{j = 1}^{n_i}
\left[  F^*(t_{ij}|x_0, \nu, \kappa_{i}, \sigma_{im_i})
	\right]^{\delta_{ij}} 
\times
\left[ 1 - F \left(t_{ij} + \frac{1}{2} \Big\vert 
		x_0, \nu, \kappa_{i}, \sigma_{im_i} \right) 
	\right]^{(1 - \delta_{ij})},
\end{align}
where
\begin{equation*}
	F^*(t_{ij}|x_0, \nu, \kappa_{i}, \sigma_{im_i})  =
    	F \left(t_{ij} + \frac{1}{2} \Big\vert x_0, \nu, \kappa_{i}, \sigma_{im_i} \right) -
    	F\left(t_{ij} - \frac{1}{2} \Big\vert x_0, \nu, \kappa_{i}, \sigma_{im_i} \right).
\end{equation*}

Under the Bayesian framework, prior distributions for the model
parameters need to be specified. For regression coefficients
$\bm{\alpha}$ and $\bm{\beta}_k$, we assign noninformative normal
priors with zero mean and large variance.
Inverse-Gamma priors are specified for the normal variances of the
frailties $\theta_{1k}$ and $\theta_2$. Dirichlet distribution is
assigned as prior for mixture weights. The priors are summarized as
follows:
\begin{align}
\label{eq:priors}
\bm{\rho} = (\rho_{1}, \rho_{2})  & \sim \mbox{Dirichlet}(\bm{\xi}), \notag \\
\alpha_l &\sim N(0, \sigma^2_{\alpha}), \quad  l = 0, \ldots, p, \notag\\
\beta_{kl} &\sim N(0, \sigma^2_{\beta}), \quad l = 0, \ldots, p, \quad k = 1, 2\\
\theta_{1k} &\sim \mbox{Inv-Gamma}(a, b), \quad k = 1, 2, \notag\\
\theta_{2} &\sim \mbox{Inv-Gamma}(a, b),   \notag
\end{align}
where $\mbox{Dirichlet}(\bm{\xi})$ is the Dirichlet distribution with
concentration parameters vector $\bm{\xi}$, $N(0, \sigma^2_{\alpha})$ and 
$N(0, \sigma^2_{\beta})$ are the normal distributions with mean zero and variance
$\sigma^2_{\alpha} > 0$ and $\sigma^2_{\beta} > 0$, respectively, and
$\mbox{Inv-Gamma}(a, b)$ is the inverse gamma distribution with shape~$a > 0$
and scale~$b > 0$. In our data analysis and simulation studies, the
hyper-parameters were set to be $\sigma^2_{\alpha} = \sigma^2_{\beta} = 10^2$,
$a = b = 1$, and $\bm{\xi} = (1, 1)$.

To derive the joint posterior distribution, let
$\bm{\Gamma} = (\bm{\Omega}, \bm{\rho}, \theta_{11}, \theta_{12}, \theta_{2})^{\top}$
denote the collection of model parameters.
With $\bm{z} = (\bm{z}_1, \cdots, \bm{z}_n)^{\top}$ and 
$\bm{m} = (m_1, \cdots, m_n)^{\top}$, 
the joint posterior density is obtained by combining the
likelihood function and the prior distributions, 
\begin{align*}
\pi( \bm{\Gamma}, \bm{z}, \bm{m} \mid  \bm{D}) 
& \propto
	\left[ \prod_{i=1}^{n} 
		L_{i} ( \bm{\Omega} | \bm{D}_i, \bm{z}_i, m_i) 
		g(\bm{z}_i|\theta_{1m_i}, \theta_2, m_i) g(m_i| \bm{\rho}) 
	\right] \\
& \quad \times
	q(\bm{\alpha})  q(\bm{\beta}_{1})  q(\bm{\beta}_{2})  
	q(\bm{\rho}) 
	q(\theta_{11}) q(\theta_{12}) q(\theta_2),
\end{align*}
where $L_{i} ( \bm{\Omega} | \bm{D}_i, \bm{z}_i, m_i)$ 
is the conditional likelihood function for subject~$i$ given
in~\eqref{eq:intvcondlklksubj}, 
$g(\bm{z}_i|\theta_{1m_i}, \theta_2, m_i)$ is the bivariate
independent normal density of the frailties of subject~$i$ 
given membership~$m_i$, 
$g(m_i| \bm{\rho})$ is the probability mass function 
of categorical distribution with probability vector $\bm{\rho}$,
and $q(\cdot)$ denotes a generic density function of its argument
and all the $q(\cdot)$'s are priors given in~\eqref{eq:priors}.
Since all the priors are proper, the posterior is proper.

We employ MCMC to make inference about the parameters. Since the FHT
density of the reflected Brownian motion does not have any standard
form, we apply the Metropolis-Hastings random walk algorithm to sample
from the full conditional distributions of all the parameters. The
implementation could be done through a generic \texttt{R} package
\texttt{NIMBLE} \citep{devalpine2017programming}. 
The model implementation details are presented in Section~5 of the
Supplementary Materials, 
which also briefly discusses how the label switching issue is handled
by imposing ordering constraint on the parameters during MCMC
using \texttt{NIMBLE}. This constraint-based approach avoids the need
for post-processing relabeling algorithms, for which methods such as
 \citet{stephens2000dealing} and implementations summarized in
  \citet{papastamoulis2016label} are widely used.
Due to the considerable number of unobserved frailties and
membership, the resulting chains exhibit
strongly auto-correlation. A sufficient number of posterior draws can
be retained after thinning the long chains, which also helps reduce
the computational burden associated with calculating the model
comparison criteria described in a later section.

\subsection{Reduced models and model selection criteria}

\begin{table}[t]
\centering
\caption{Summary of candidate models. 
CS-C-FV: component-specific coefficients and frailty variances model;
CS-I-FV: component-specific intercept and frailty variances model;
CS-I: component-specific intercept model;
CS-N: none component specific (single component frailty model)
}  
\label{tab:modelsumm}
\begin{tabular}{lll}
\toprule
Model & Volatility $\sigma_{ik}$ & Constraint\\
\midrule
CS-C-FV
& $\log(\sigma_{ik}) = \boldsymbol{X}_i^{\top} \bm{\beta_k} + Z_{1ik}$ 
&  \\
& $Z_{1ik} \sim N(0, \theta_{1k})$ & \\
[0.8em]
CS-I-FV
& $\log(\sigma_{ik}) = \beta_{k0} + \sum_{l=1}^p X_{il} \beta_l + Z_{1ik}$ 
& $\beta_{l} = \beta_{1l} = \beta_{2l}$ for $l = 1, \cdots, p$\\
& $Z_{1ik} \sim N(0, \theta_{1k})$ & \\
[0.8em]
CS-I
& $\log(\sigma_{ik}) = \beta_{k0} + \sum_{l=1}^p X_{il} \beta_l + Z_{1ik}$ 
& $\beta_{l} = \beta_{1l} = \beta_{2l}$ for $l = 1, \cdots, p$, \\ 
& $Z_{1ik} \sim N(0, \theta_{11})$ 
& and $\theta_{11} = \theta_{12}$ \\
[0.8em]
CS-N
& $\log(\sigma_{i}) = \beta_{0} + \sum_{l=1}^p X_{il} \beta_l + Z_{1i}$ 
& Only one component \\ 
& $Z_{1i} \sim N(0, \theta_{1})$ 
&  \\
\bottomrule
\end{tabular}
\end{table}

Given the setup in~\eqref{eq:mixmodelpara} as the full model, several
reduced models are proposed. Model~2 is a reduced model of the full
model, imposing the restriction that both components share identical
regression coefficients, i.e., $\beta_{1l} = \beta_{2l}$ for $l = 1,
\cdots, p$. This assumption suggests that covariates have the same
impact across different population subgroups. Model~3 further
simplifies Model~2 by constraining the frailty variances of the two
components to be identical, denoted as $\theta_{11} = \theta_{12}$.
This assumption indicates that the frailties of both components are
expected to be from the same distribution. For ease of reference, we
classify the models based on whether the covariate coefficients or
frailty variances are component-specific (CS). Accordingly, we label
the full and the two reduced models as the CS-C-FV (component-specific
coefficients and frailty variances) model, the CS-I-FV
(component-specific intercept and frailty variances) model, and the
CS-I (component-specific intercept) model, respectively. The 
independent-frailty model proposed 
by \citet{xie2025recurrent} includes only single component is referred 
as CS-N (none component specific) model here. 
A comparative summary of these models is presented in
Table~\ref{tab:modelsumm}.

Two model comparison criteria, deviance information criterion (DIC) 
\citep{spiegelhalter2002bayesian} and logarithm of the pseudo-marginal 
likelihood (LPML) \citep{geisser1979predictive, gelfand1994bayesian}
are considered to conduct model selection. For a candidate model,
define deviance
\begin{equation*}
	\mbox{Dev} (\bm{\Gamma}) = -2 \sum_{i=1}^n \log 
	L_{i, obs} (\bm{\Gamma} | \bm{D}_i),
\end{equation*}
where $L_{i, obs} (\bm{\Gamma} | \bm{D}_i)$ is the contribution to
the observed likelihood from subject~$i$.
The DIC is then defined as
\begin{equation*}
	\mbox{DIC} = \mbox{Dev}( \bar{\bm{\Gamma}} ) + 2p_D,
\end{equation*}
where 
$p_D = \overline{\mbox{Dev}}(\bm{\Gamma}) - \mbox{Dev}(\bar{\bm{\Gamma}})$ is 
the effective number of model parameters, and $\bar{\bm{\Gamma}}$ and
$\overline{\mbox{Dev}}(\bm{\Gamma})$ are the posterior means of
$\bm{\Gamma}$ and $\mbox{Dev}(\bm{\Gamma})$, respectively.
A lower value of DIC means a better model.

As there is no closed form for the observed data likelihood which 
integrates out all the latent variables, 
Monte Carlo approximation is used to approximate the integral. 
The observed data likelihood of subject~$i$ has the form
\begin{align*} 
L_{i, obs} (\bm{\Gamma} | \bm{D}_i) 
	&= \sum_{k=1}^2 g(m_i = k|\bm{\rho}) 
			\int L_{i} (\bm{\Omega} | \bm{D}_i, \bm{z}_i, m_i = k)
			g(\bm{z}_i|\theta_{1m_i}, \theta_2, m_i = k) \dd \bm{z}_i , 
\end{align*}
where $\bm{\Gamma}$ is the set of model parameters and $L_{i}
(\bm{\Omega} | \bm{D}_i, \bm{z}_i, m_i)$ is given
in~\eqref{eq:intvcondlklksubj}. 
The random effects $\bm{z}_i$ are continuous, and the corresponding
integral is approximated using Monte Carlo integration. In contrast,
the latent membership $m_i$ is a discrete random variable taking
values in $\{1,2\}$, and integration with respect to $m_i$ is
performed by summation over its possible values. Then the observed
data likelihood for subject~$i$ can be derived by
\begin{align}
\label{eq:MCapprox}
L_{i, obs} (\bm{\Gamma} | \bm{D}_i)  
\approx \sum_{k=1}^2 g(m_i = k|\bm{\rho}) 
	\left[ \frac{1}{S} \sum_{s=1}^S 
	L_{i} (\bm{\Omega} | \bm{D}_i, \bm{z}_i^{(s)}, m_i = k) \right],
\end{align}
where $\bm{z}_i^{(1)}, \cdots, \bm{z}_i^{(S)}$ are Monte Carlo samples
that each can be generated independently from a bivariate normal
distributions $N(\bm{0}, \bm{\Sigma})$, where the covariance matrix
$\bm{\Sigma} = \mathrm{diag}(\theta_{1m_i}, \theta_2)$ conditional on
$m_i$, and $S$ is the Monte Carlo sample size.

LPML is calculated based on conditional predictive ordinate (CPO). Let
$D_{-i} = \{(\bm{t}_j, \bm{\delta}_j, X_j): j = 1,\ldots, n; j \ne i\}$, 
denote the observed data without the $i$th~subject. The CPO for
the $i$th~subject is the leave-one-out predictive likelihood
\begin{align*}
\mbox{CPO}_i = 
	\int L_{i, obs} (\bm{\Gamma}|\bm{D}_i) 
		q(\bm{\Gamma}|D_{-i}) \dd \bm{\Gamma},
\end{align*}
where
$q(\bm{\Gamma}|D_{-i})$ is the marginal posterior distribution
of $\bm{\Gamma}$ with the $i$th subject deleted. The Monte
Carlo estimate of $\mbox{CPO}_i$ \citep{dey1997bayesian} is given by
\begin{equation}
  \label{eq:cpo}
   \widehat{\mbox{CPO}}_{i} = \left[\frac{1}{U}\sum_{u=1}^U 
   \frac{1}{L_{i, obs} (\bm{\Gamma}^{(u)}|\bm{D}_i)}\right]^{-1},
\end{equation}
where
$\bm{\Gamma}^{(u)} = (\bm{\alpha}^{(u)}, \bm{\beta}_1^{(u)}, \bm{\beta}_2^{(u)}, \bm{\rho}^{(u)},\theta_{11}^{(u)}, \theta_{12}^{(u)},\theta_{2}^{(u)})$, 
for $u = 1, \ldots, U$, denotes the $u$th MCMC draw from the posterior
distribution, and $U$ is the total number of posterior draws.
Each term $L_{i, obs}$ in~\eqref{eq:cpo} can be approximated the same
way as in~\eqref{eq:MCapprox}. Then LPML can be calculated by
\begin{equation*}
  \widehat{\mbox{LPML}} = \sum_{i = 1}^n \log (\widehat{\mbox{CPO}}_i).
\end{equation*}
Models with higher LPML are preferred.

The calculation of model comparison criteria is based on the observed
data likelihood, which requires integrating all the random effects and
latent memberships out.  This process is computationally intensive. To
reduce the computational burden, we used a relatively small number of
posterior samples to calculate the model comparison
criteria in our simulation study.

\section{Results} \label{sec:aplc}

\begin{table} 
\caption{
Model comparison for the four models fitted to the motivating data. A
lower value of DIC (deviance information criterion) and a higher value
of LPML (logarithm of the pseudo-marginal likelihood) indicate a
better model.
}
\label{tab:aplcModCpr}
\begin{center}
\begin{tabular}{lrrrr}
	\toprule
	& CS-C-FV  
	& CS-I-FV     
	& CS-I        
	& CS-N \\
	\midrule
	DIC     &     130843.3  &   130871.8   &   130947.3  &   131344.0 \\
	LPML    &   $-$65444.2  & $-$65449.2   & $-$65498.5  & $-$65676.7 \\
	\bottomrule
\end{tabular}
\end{center}
\end{table}

Three proposed models were applied to analyze the motivating
data. The priors for all the parameters were specified as shown in
Section~\ref{subsec:lklh}. The lower boundary~$\nu$ was set to $3.9$
mmol/l ($70$ mg/dl), which is the clinical standard for hypoglycemic
events \citep{seaquist2013hypoglycemia}. The starting point~$x_0$ of
the Brownian motion after each hypoglycemic event was set to be $10$,
which is the rounded integer of the median of the baseline fasting
glucose level of all patients. 
For each model, an MCMC was run for $150,000$ iterations and thinned
by $10$ after discarding the first $30,000$ iterations as burn-in.
Sensitivity analysis about the alternative starting values of some
parameters are presented in Supplementary Materials Section~4.2.
The
convergence of the MCMC chains was monitored by traceplots, which are
shown in the Section~2 of the Supplementary Materials. The results of
DIC and LPML for the three models are presented in
Table~\ref{tab:aplcModCpr}, alongside the results for the
independent-frailty model (referred as CS-N in this study) as reported
by \citet{xie2025recurrent}. 
Both DIC and LPML favor the full model, CS-C-FV model, which suggests
that component-specific regression coefficients and frailty variances
are crucial for capturing differing covariate effects and unobserved
heterogeneity, respectively between the two subgroups.

\begin{sidewaystable} [tbp]
\centering
\caption{Estimated parameters of the CS-C-FV model. 
	BMI, body mass index; 
	BP, blood pressure;
	SD, standard deviation; 
	CI, $95$\% HPD credible interval or $95$\% confident interval.}
\label{tab:aplcEst}
\begin{tabular}{lrrcrrcrrc}
\toprule
				& \multicolumn{6}{c}{Volatility} 
				& \multicolumn{3}{c}{Upper reflection barrier} \\
				\cmidrule(lr){2-7}   \cmidrule(lr){8-10}
				& \multicolumn{3}{c}{Group 1} 
				& \multicolumn{3}{c}{Group 2} & \\
				\cmidrule(lr){2-4}   \cmidrule(lr){5-7}  
Covariates  	& Mean & SD & 95\% CI 
                & Mean & SD & 95\% CI 
                & Mean & SD & 95\% CI \\ 
\midrule
Intercept           & $-$0.171 & 0.130 & [$-$0.411, 0.072] 
					& 1.410 & 0.027 & \textbf{[1.357, 1.458]} 
					& 2.839 & 0.062 & \textbf{[2.715, 2.955]} \\ 
Fasting glucose 	& $-$0.139 & 0.051 & \textbf{[$-$0.24, $-$0.04]} 
					& $-$0.016 & 0.013 & [$-$0.041, 0.009] 
					& 0.11 & 0.031 & \textbf{[0.051, 0.17]} \\ 
Adiponectin         & 0.134 & 0.066 & \textbf{[0.016, 0.267]} 
					& 0.019 & 0.012 & [$-$0.004, 0.043] 
					& 0.034 & 0.03 & [$-$0.026, 0.09] \\ 
Fasting insulin 	& $-$0.272 & 0.056 & \textbf{[$-$0.381, $-$0.159]} 
					& $-$0.036 & 0.014 & \textbf{[$-$0.062, $-$0.008]} 
					& 0.164 & 0.032 & \textbf{[0.096, 0.222]} \\ 
Height 				& 0.067 & 0.050 & [$-$0.024, 0.174] 
					& $-$0.049 & 0.013 & \textbf{[$-$0.072, $-$0.023]} 
					& 0.019 & 0.03 & [$-$0.035, 0.081] \\ 
BMI 				& $-$0.092 & 0.061 & [$-$0.209, 0.024] 
					& $-$0.046 & 0.012 & \textbf{[$-$0.069, $-$0.02]} 
					& -0.11 & 0.033 & \textbf{[$-$0.17, $-$0.043]} \\ 
Diastolic BP 		& $-$0.064 & 0.062 & [$-$0.188, 0.049] 
					& $-$0.023 & 0.015 & [$-$0.052, 0.005] 
					& 0.061 & 0.034 & [$-$0.005, 0.128] \\ 
Systolic BP 		& 0.053 & 0.062 & [$-$0.068, 0.176] 
					& 0.007 & 0.013 & [$-$0.019, 0.032] 
					& -0.023 & 0.031 & [$-$0.08, 0.042] \\ 
Heart rate 			& 0.082 & 0.062 & [$-$0.03, 0.204] 
					& 0.000 & 0.012 & [$-$0.022, 0.023] 
					& 0.05 & 0.028 & [$-$0.002, 0.105] \\ 
Duration diabetes 	& 0.190 & 0.047 & \textbf{[0.092, 0.274]} 
					& 0.020 & 0.011 & [$-$0.002, 0.044] 
					& -0.043 & 0.027 & [$-$0.091, 0.011] \\ 
LM 					& 0.177 & 0.109 & [$-$0.021, 0.385] 
					& 0.092 & 0.023 & \textbf{[0.047, 0.137]} 
					& -0.124 & 0.056 & \textbf{[$-$0.24, $-$0.027]} \\ 
tzd-only 			& $-$0.445 & 0.200 & \textbf{[$-$0.828, $-$0.073]} 
					& $-$0.166 & 0.060 & \textbf{[$-$0.282, $-$0.054]} 
					& 0.396 & 0.144 & \textbf{[0.129, 0.682]} \\ 
sulf-only 			& 0.058 & 0.120 & [$-$0.187, 0.277] 
					& $-$0.015 & 0.027 & [$-$0.063, 0.044] 
					& 0.067 & 0.065 & [$-$0.071, 0.188] \\ 
[1em]
Frailty Std 		& 0.773 & 0.051 & \textbf{[0.674, 0.867]} 
					& 0.249 & 0.012 & \textbf{[0.226, 0.274]} 
					& 0.699 & 0.027 & \textbf{[0.647, 0.751]} \\ 
Mixture_Weight 		& 0.440 & 0.023 & \textbf{[0.399, 0.486]} 
					& 0.560 & 0.023 & \textbf{[0.514, 0.601]} 
					&   &   &  \\ 
\bottomrule
\end{tabular}
\end{sidewaystable}

\begin{figure} [tbp]
	\centering
	\includegraphics[width=\linewidth]{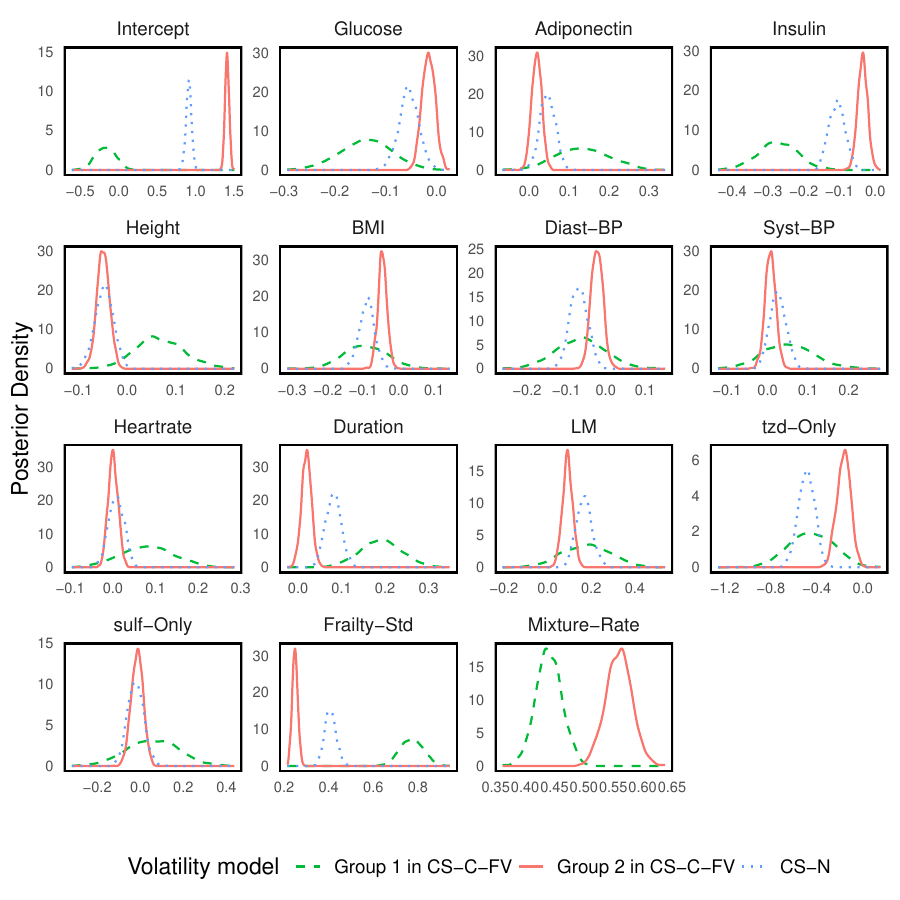}
	\caption{Posterior density of estimates in volatility model in 
	CS-C-FV and CS-N model.}	
	\label{fig:sigmaPostDistCpr}
\end{figure}

The posterior density for the regression coefficients related to
volatility of the CS-C-FV model and CS-N model are depicted in
Figure~\ref{fig:sigmaPostDistCpr}. Additionally, the point estimate,
standard error, and $95\%$ highest posterior density (HPD) credible
interval for all parameters of the CS-C-FV model are summarized in
Table~\ref{tab:aplcEst}. From Table~\ref{tab:aplcEst}, we generally
find that the covariate effects in the upper reflection barrier
estimated by the CS-C-FV model are similar to those reported for the CS-N
model by \citet{xie2025recurrent}. This similarity is anticipated
since the mixture structure is considered solely in
volatility. A more detailed examination of the covariate effects in
the volatility model, as depicted in
Figure~\ref{fig:sigmaPostDistCpr}, reveals that the estimates from the
CS-N model fall between the two groups from the CS-C-FV model. 
This indicates that the mixture structure offers richer and more informative characterization.  Notably, all
coefficients that are significant in the volatility model of the CS-N
model remain significant in either group~1 or~2, with an exception of
the baseline diastolic blood pressure. 
The comparison of the volatility intercept terms between the two
groups indicates a significant difference.  As mentioned in
Section~\ref{subsec:modformula}, in general, larger volatility is
associated with an increased risk of hypoglycemic events.  Hence, the
significant difference in volatility intercept terms between the two
groups justifies categorizing them as low risk (group 1) or high risk
(group 2) for hypoglycemia, with a smaller intercept indicating lower
risk and a larger intercept indicating higher risk. According to the
estimates of the mixture weight, the proportions of the low-risk and
high-risk groups are approximately $44\%$ and $56\%$, respectively.

Among the low-risk group, patients with lower baseline fasting
glucose, lower baseline fasting insulin, higher baseline adiponectin,
longer duration of diabetes are associated with larger volatility so 
higher risk of hypoglycemic events. For the patients in the high-risk
group, lower baseline fasting insulin, lower height, lower BMI
are associated with a higher risk of hypoglycemia. High-risk patients
who received LM$75/25$ appear to have higher volatility or  a higher risk
of hypoglycemia compared to those who received glargine. For oral
antihyperglycemic drugs, regardless of the group, patients solely
received thiazolidinedione appear to exhibit reduced volatility or a
lower risk of hypoglycemia when compared to those received both
thiazolidinedione and sulfonylurea. There is no significant difference
in patients received solely with sulfonylureas compared to those
receiving both oral antihyperglycemic medications.

\begin{figure} [tbp]
\centering
\includegraphics[width=\linewidth]{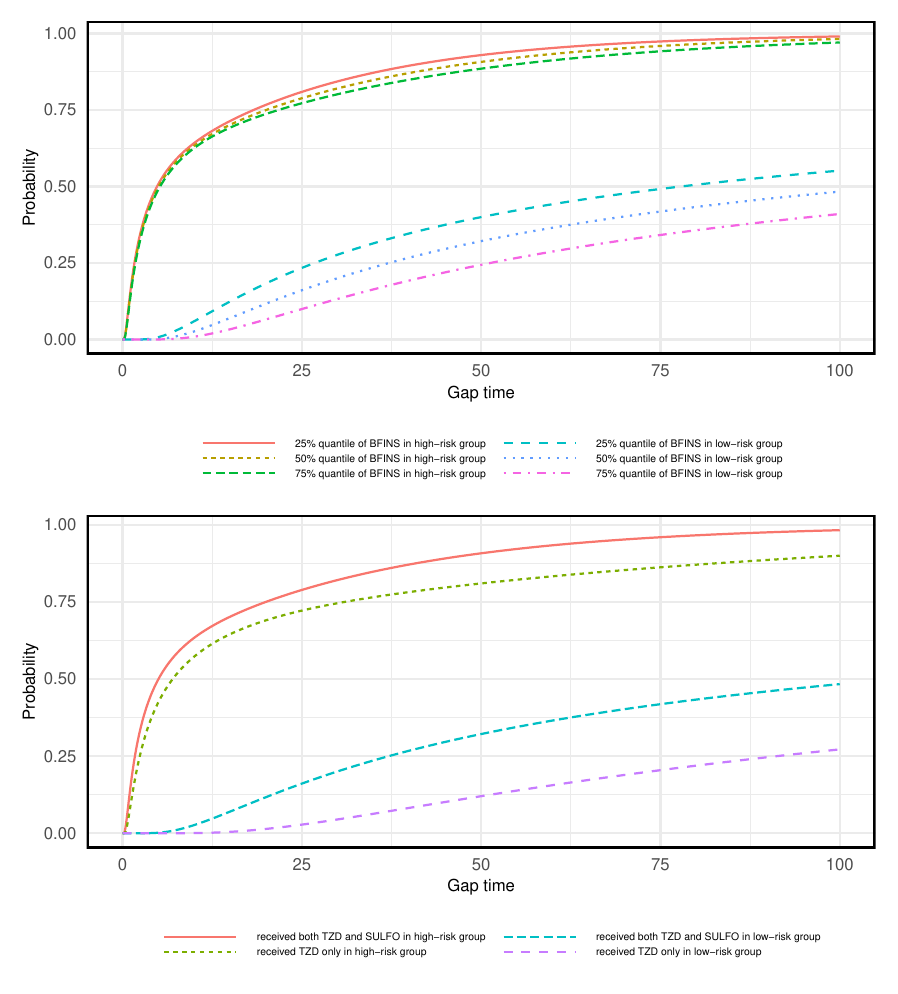}
\caption{Predictive distribution functions were derived using the
estimated parameters at different covariate levels by risk group.
Three levels corresponding to the $25\%$, $50\%$, and $75\%$ quantiles
of baseline fasting insulin and two levels of \textsf{tzd-only} (0 and
1) were considered, shown in the top and bottom panels, respectively.
BFINS: baseline fasting insulin; TZD: thiazolidinedione; SULFO:
sulfonylurea.
}	
\label{fig:predPlot}
\end{figure}

For covariates significantly impacting both the low-risk and high-risk
groups, baseline fasting insulin and \textsf{tzd-only}, we have
included plots of the predictive distribution derived from the fitted
parameters for the two groups in Figure~\ref{fig:predPlot}. These
illustrations demonstrate the influence of baseline insulin levels and
the exclusive use of thiazolidinediones on the distribution across the
two groups. The figure indicates that when controlling other
covariates at the same level, the median predictive gap time for
patients within the high-risk group is substantially less compared to
those in the low-risk group. This suggests a higher frequency of
hypoglycemic events among high-risk patients. Additionally, the plot
also highlights the distinct effects of various covariates on the two
groups. As previously mentioned, for example, receiving both
medications is associated with an increased risk of hypoglycemia when
compared to only receiving thiazolidinedione. Upon further examination
of the impact on different groups, it is observed that taking both
medications significantly reduces the gap time for patients in the
low-risk group more than in the high-risk group, indicating a markedly
increased risk of hypoglycemia.

\begin{figure} [tbp]
\centering
\includegraphics[width=\linewidth]{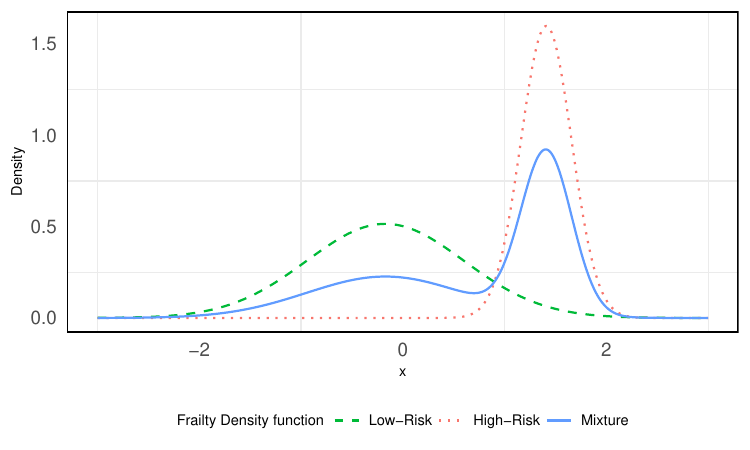}
\caption{Frailty density functions for the low- and high-risk groups
and the corresponding mixture frailty density function based on the
estimated parameters.}	
\label{fig:mixFrailtyDensMeanIntercept}
\end{figure}

When we examine the variability in frailty across the two identified
patient groups, a notable distinction emerges in  frailty
standard deviation. The frailty distributions for each group and the 
mixture frailty distribution are shown in 
Figure~\ref{fig:mixFrailtyDensMeanIntercept}. Specifically, the 
high-risk group is characterized
by a relatively smaller standard deviation in frailty, suggesting a
more homogenous reaction to hypoglycemic risk factors within this
group. Conversely, the standard deviation is significantly larger in
the low-risk group, pointing to a higher degree of between-subject
variability. This greater variance within the low-risk group indicates
that, despite being categorized as less susceptible to hypoglycemic
events, there is a wide range of underlying susceptibilities and
responses to treatment. This observation underscores the complexity of
managing diabetes and highlights the importance of personalized
treatment plans that consider the differences in patient
responses, even within the same risk group.

In the upper reflection barrier model, patients with lower baseline
fasting glucose, lower fasting insulin, higher BMI are associated
lower upper reflection barrier or a higher risk of hypoglycemia. For the
oral antihyperglycemic drugs, the patients only received
thiazolidinedione appear to have a lower risk of hypoglycemia compared
to those who received both two oral antihyperglycemic. Patients who
received LM$75/25$ appear to have a lower reflection barrier or higher
risk of hypoglycemia compared to those who received glargine. These
results are similar to those reported in the earlier study.

\section{Simulation} \label{sec:simulation}
Simulation studies were conducted to assess the estimation procedure
and model selection criteria. Two covariates were considered for each
subject~$i$: $X_{i1}$, baseline insulin level, and $X_{i2}$, baseline
body mass index (BMI). To mimic the covariate structure in the
motivating data, a bivariate gamma distribution with a normal copula
was fitted using the \proglang{R} package \pkg{copula}
\citep{hofert2018elements}, and the two covariates were then generated
from the fitted distribution.

Three mixture FHT models in Table~\ref{tab:modelsumm} were used to
generate recurrent event times with covariate 
vector~$\bm{X_i} = (1, X_{i1}, X_{i2})^{\top}$. The true values of
parameters in the full model were set with
$\bm{\alpha} = (2.9, 0.2, -0.1)^{\top}$,
$\bm{\beta}_1 = (-0.2, -0.3, -0.1)^{\top}$,  
$\bm{\beta}_2 = (1.4, -0.05, -0.05)^{\top}$, and 
$\bm{\rho} = (0.4, 0.6)^{\top}$. The frailties $z_{1i1}$ and $z_{1i2}$
in volatility for two groups were generated from the normal
distribution with mean~zero and variance $\theta_{11} = 0.5$ and
$\theta_{12} = 0.1$. The frailty $z_{2i}$ in the upper reflection
barrier was generated from the normal distribution mean~zero and
variance $\theta_{2} =  0.3$. The CS-I model constraints the regression
coefficients in volatility of two components to be 
$\bm{\beta}_1 = (-0.2, -0.05, -0.05)^{\top}$
and $\bm{\beta}_2 = (1.4, -0.05, -0.05)^{\top}$, i.e., 
$\beta_{11} = \beta_{21} = -0.05$ and $\beta_{12} = \beta_{22} = -0.05$.
The CS-N model does not involve a mixture structure (simulation
studies of the CS-N model have been reported in
\citet{xie2025recurrent} and are not presented here). The starting
point~$x_0$ and the lower boundary~$\nu$ were set to be $x_0 = 10$ and
$\nu = 3.9$, respectively.

For each model, event gap times were simulated from the FHT
distribution using the rejection sampling algorithm of
\citet{xie2025recurrent}. The gap times were generated until the
accumulated gap times reached or exceeded the follow-up time for
subject~$i$. Follow-up times were independently drawn from the
empirical distribution observed in the motivating data. 
The group membership of each subject was then generated from a
categorical distribution with the specified mixture weights.
Different levels of sample sizes, $n \in \{200, 500, 1000, 2000\}$
were considered. For each simulation setting, $100$~datasets were
generated.

For each dataset, Bayesian inference was performed using the prior
distributions defined in Equation~\eqref{eq:priors}, with
hyperparameters described in Section~\ref{subsec:lklh}. For each
dataset, an MCMC chain of 100{,}000 iterations was run, with the first
50{,}000 discarded as burn-in and the remaining iterations thinned by
a factor of 100. Convergence was assessed using the
Heidelberger--Welch diagnostic \citep{heidelberger1983simulation}
implemented in the \texttt{R} package \texttt{CODA}
\citep{plummer2006coda}. Posterior summaries and model selection
criteria were computed from the resulting 500 MCMC samples.

\begin{table} [tbp]
\caption{Results of parameter estimation under correct specifications
	for three models with sample size $n \in \{1000, 2000\}$. 
	SD, posterior standard deviation; 
	ESD, empirical standard deviation; 
	CR, coverage rate of $95\%$ HPD credible interval.}  
\label{tab:simuResSum}
\centering
\begin{tabular}{crrrrrrrrrrr}		
	\toprule
			&  &  & \multicolumn{4}{c}{$n = 1000$} & \multicolumn{4}{c}{$n = 2000$} \\
				\cmidrule(lr){4-7} \cmidrule(lr){8-11}
	Model & Para  & True & Bias & SD & ESD & CR & Bias & SD & ESD & CR \\ 
\midrule 
CS-C-FV
&$\alpha_0$ & 2.90 & 0.009 & 0.039 & 0.035 & 0.98 & $-$0.001 & 0.027 & 0.025 & 0.98 \\ 
&$\alpha_1$ & 0.20 & 0.003 & 0.044 & 0.043 & 0.96 & 0.001 & 0.031 & 0.027 & 0.97 \\ 
&$\alpha_2$ & $-$0.10 & 0.006 & 0.038 & 0.033 & 0.97 & 0.008 & 0.027 & 0.026 & 0.93 \\ 
&$\beta_{10}$ & $-$0.20 & $-$0.055 & 0.117 & 0.107 & 0.95 & $-$0.050 & 0.083 & 0.080 & 0.93 \\ 
&$\beta_{11}$ & $-$0.30 & $-$0.016 & 0.099 & 0.091 & 0.95 & $-$0.005 & 0.067 & 0.071 & 0.92 \\ 
&$\beta_{12}$ & $-$0.10 & $-$0.003 & 0.087 & 0.081 & 0.97 & $-$0.002 & 0.060 & 0.062 & 0.90 \\ 
&$\beta_{20}$ & 1.40 & $-$0.018 & 0.027 & 0.026 & 0.91 & $-$0.013 & 0.020 & 0.019 & 0.90 \\ 
&$\beta_{21}$ & $-$0.05 & 0.001 & 0.022 & 0.019 & 0.96 & $-$0.000 & 0.015 & 0.014 & 0.96 \\ 
&$\beta_{22}$ & $-$0.05 & $-$0.001 & 0.021 & 0.018 & 0.98 & $-$0.001 & 0.014 & 0.016 & 0.96 \\ 
&$\rho_1$ & 0.40 & $-$0.012 & 0.034 & 0.030 & 0.96 & $-$0.012 & 0.026 & 0.024 & 0.90 \\ 
&$\rho_2$ & 0.60 & 0.012 & 0.034 & 0.030 & 0.96 & 0.012 & 0.026 & 0.024 & 0.90 \\ 
&$\theta_{11}$ & 0.50 & 0.030 & 0.143 & 0.095 & 0.98 & $-$0.011 & 0.110 & 0.081 & 0.98 \\ 
&$\theta_{12}$ & 0.10 & 0.013 & 0.014 & 0.012 & 0.89 & 0.008 & 0.010 & 0.008 & 0.88 \\ 
&$\theta_2$ & 0.30 & 0.018 & 0.040 & 0.034 & 0.98 & 0.007 & 0.027 & 0.027 & 0.93 \\ 
\midrule
CS-I-FV
&$\alpha_0$ & 2.90 & 0.010 & 0.039 & 0.035 & 0.99 & $-$0.001 & 0.027 & 0.024 & 0.97 \\
&$\alpha_1$ & 0.20 & 0.003 & 0.044 & 0.046 & 0.95 & $-$0.001 & 0.031 & 0.028 & 0.95 \\ 
&$\alpha_2$ & $-$0.10 & 0.004 & 0.038 & 0.035 & 0.98 & 0.005 & 0.027 & 0.027 & 0.95 \\ 
&$\beta_{10}$ & $-$0.20 & $-$0.056 & 0.118 & 0.086 & 0.96 & $-$0.035 & 0.087 & 0.075 & 0.96 \\ 
&$\beta_{20}$ & 1.40 & $-$0.019 & 0.028 & 0.026 & 0.94 & $-$0.012 & 0.021 & 0.019 & 0.97 \\ 
&$\beta_1$ & $-$0.05 & $-$0.002 & 0.021 & 0.019 & 0.98 & $-$0.001 & 0.015 & 0.014 & 0.94 \\ 
&$\beta_2$ & $-$0.05 & $-$0.002 & 0.021 & 0.019 & 0.97 & $-$0.002 & 0.014 & 0.015 & 0.95 \\ 
&$\rho_1$ & 0.40 & $-$0.015 & 0.036 & 0.027 & 0.96 & $-$0.008 & 0.028 & 0.025 & 0.97 \\ 
&$\rho_2$ & 0.60 & 0.015 & 0.036 & 0.027 & 0.96 & 0.008 & 0.028 & 0.025 & 0.97 \\ 
&$\theta_{11}$ & 0.50 & 0.016 & 0.153 & 0.087 & 0.99 & $-$0.006 & 0.119 & 0.085 & 0.97 \\ 
&$\theta_{12}$ & 0.10 & 0.014 & 0.014 & 0.012 & 0.90 & 0.007 & 0.010 & 0.009 & 0.91 \\ 
&$\theta_2$ & 0.30 & 0.019 & 0.040 & 0.036 & 0.97 & 0.007 & 0.027 & 0.028 & 0.91 \\ 
\midrule
CS-I
&$\alpha_0$ & 2.90 & 0.007 & 0.040 & 0.036 & 0.98 & 0.002 & 0.028 & 0.027 & 0.93 \\ 
&$\alpha_1$ & 0.20 & 0.003 & 0.045 & 0.046 & 0.94 & 0.001 & 0.031 & 0.027 & 0.98 \\ 
&$\alpha_2$ & $-$0.10 & 0.005 & 0.039 & 0.035 & 0.97 & 0.006 & 0.027 & 0.028 & 0.91 \\ 
&$\beta_{10}$ & $-$0.20 & $-$0.012 & 0.056 & 0.096 & 0.99 & $-$0.009 & 0.037 & 0.036 & 0.96 \\ 
&$\beta_{20}$ & 1.40 & $-$0.008 & 0.034 & 0.022 & 0.93 & $-$0.005 & 0.015 & 0.014 & 0.93 \\
&$\beta_1$ & $-$0.05 & $-$0.003 & 0.021 & 0.020 & 0.94 & $-$0.001 & 0.014 & 0.013 & 0.95 \\ 
&$\beta_2$ & $-$0.05 & $-$0.000 & 0.020 & 0.018 & 0.97 & $-$0.002 & 0.014 & 0.015 & 0.92 \\ 
&$\rho_1$ & 0.40 & $-$0.002 & 0.022 & 0.040 & 0.95 & $-$0.002 & 0.014 & 0.014 & 0.92 \\ 
&$\rho_2$ & 0.60 & 0.002 & 0.022 & 0.040 & 0.95 & 0.002 & 0.014 & 0.014 & 0.92 \\ 
&$\theta_{1}$ & 0.10 & 0.013 & 0.012 & 0.037 & 0.87 & 0.005 & 0.007 & 0.007 & 0.89 \\ 
&$\theta_{2}$ & 0.30 & 0.018 & 0.040 & 0.035 & 0.97 & 0.007 & 0.028 & 0.029 & 0.90 \\
\bottomrule
\end{tabular}
\end{table}

Table~\ref{tab:simuResSum} summarizes the estimation results for the
three candidate models when the data were generated under the correct
model specification. The bias, defined as the difference between the
true parameter value and its point estimate (the posterior mean), is
close to zero for all parameters at both sample sizes. The average
posterior standard deviation closely matches the empirical standard
deviation of the estimates, except for the frailty variance
$\theta_{11}$ or $\theta_1$ across models; however, this agreement
improves as the sample size increases from $1000$ to $2000$. The
empirical coverage probabilities of the $95\%$ highest posterior
density (HPD) credible intervals are close to their nominal level for
most parameters. 
Consistent performance is also observed at smaller sample sizes
($n = 200$ and $500$), with estimation accuracy and interval coverage
remaining comparable to those at larger sample sizes. These results
indicate that the estimation procedure is robust across a range of
sample size settings. Additional details are provided in
Section~4.1 of the Supplementary Materials.

\begin{table} [tbp]
\caption{Model comparison result with DIC and LPML with sample size 
$n \in \{1000 \mbox{ and } 2000\}$.
Freq (\%): frequency of the correct model being selected; 
Mean: average of the DIC or LPML.}
\label{tab:simuModCprSum}
\begin{center}
\begin{tabular}{cccrrrrrr}
\toprule
&   &  &
\multicolumn{2}{c}{CS-C-FV}  &
\multicolumn{2}{c}{CS-I-FV}  &
\multicolumn{2}{c}{CS-I} \\
\cmidrule(lr){4-5} \cmidrule(lr){6-7} \cmidrule(lr){8-9} 
True Model  & Criterion & $n$ & Freq & Mean & Freq & Mean & Freq & Mean \\ 
\midrule
CS-C-FV 
& DIC	& 200  & 54 & 12166.0 & 21 & 12167.5 & 24 & 12167.9 \\ 
&       & 500  & 75 & 29766.6 & 19 & 29770.6 & 6 & 29773.8 \\
&		& 1000  & 87 & 59511.9   & 8 & 59521.3    & 5  & 59526.6 \\
&       & 2000  & 100& 119067.9  & 0 & 119087.3   & 0  & 119096.6 \\

& LPML  & 200  & 40 & $-$6086.5 & 26 & $-$6086.6 & 35 & $-$6086.7 \\ 
&       & 500  & 66 & $-$14889.4 & 14 & $-$14891.4 & 20 & $-$14891.3 \\
&		& 1000  & 82 & $-$29762.5& 6 & $-$29766.7 & 12 & $-$29768.2 \\
&       & 2000  & 95 & $-$59540.5& 1 & $-$59549.7 & 4  & $-$59553.1 \\
[1ex]
CS-I-FV 
& DIC	& 200  & 22 & 11996.6 & 37 & 11995.3 & 40 & 11995.9 \\ 
&       & 500  & 19 & 29518.4 & 49 & 29516.8 & 31 & 29518.7 \\ 
& 		& 1000  & 15 & 58942.5  & 67 & 58993.3  & 18 & 58945.2 \\ 
&       & 2000  & 20 & 118024.3 & 68 & 118022.5 & 12 & 118031.4 \\ 
& LPML  & 200  & 14 & $-$6002.1 & 36 & $-$6000.7 & 49 & $-$6000.6 \\
&		& 500  & 14 & $-$14763.6 & 46 & $-$14762.3 & 39 & $-$14762.4 \\
&		& 1000  & 11 & $-$29477.0 & 60 & $-$29502.4 & 29 & $-$29477.5 \\
&		& 2000  & 20 & $-$59017.1 & 67 & $-$59016.4 & 13 & $-$59019.5 \\ 
[1ex]
CS-I 		
& DIC	& 200  & 16 & 11703.7 & 36 & 11702.2 & 48 & 11702.4 \\  
&       & 500  & 14 & 28701.5 & 25 & 28700.1 & 61 & 28699.3 \\ 
&		& 1000  & 2 & 57341.0 & 1 & 57338.7 & 97 & 57335.1 \\ 
&       & 2000  & 0 & 114703.4 & 3 & 114701.2 & 97 & 114697.1 \\
& LPML  & 200  & 10 & $-$5855.4 & 35 & $-$5854.0 & 55 & $-$5853.6 \\ 
&       & 500  & 14 & $-$14354.7 & 19 & $-$14354.1 & 67 & $-$14352.8 \\ 
&		& 1000  & 3 & $-$28675.6 & 5 & $-$28674.7 & 92 & $-$28672.4 \\
&       & 2000  & 1 & $-$57356.3 & 6 & $-$57355.2 & 93 & $-$57352.9 \\

\bottomrule
\end{tabular} 
\end{center}
\end{table}

We also evaluated the performance of the model comparison criteria
based on a Monte Carlo sample size~$S = 1000$ to approximate the
integral, with results summarized in Table~\ref{tab:simuModCprSum} for
sample sizes $n = 200, 500, 1000,$ and $2000$.
When either the CS-C-FV model or the CS-I model was the correct
specification, both DIC and LPML show a fair level of accuracy in
selecting the best model across all sample sizes, with selection
frequencies improving as the sample size increases.
When the CS-I-FV model was the correct specification, the frequencies
of correctly selecting the model using DIC and LPML remained the
highest among the candidates for $n \ge 500$, although they were
comparatively lower than in the other two scenarios.
For the smallest sample size $n = 200$, we observe that DIC and LPML
tend to favor the simpler CS-I model when the true model is CS-I-FV,
reflecting the limited ability of these criteria to distinguish
between highly similar models in small samples, particularly when the
models differ by only one random-effect variance parameter.

In addition to the above setting, we considered an alternative
simulation scenario motivated by the observation that the treatment
effects with opposite directions across latent components may be
masked in a homogeneous model. In this scenario, the treatment effect
was specified to be positive in one mixture component and negative in
the other. Under this setting, we found that the proposed mixture
model, CS-C-FV, was able to recover the component-specific treatment
effects, whereas the CS-N frequently failed to detect a statistically
significant treatment effect. Detailed simulation results for this
scenario are provided in the Section~4.2 of the Supplementary
Materials. Comparison of the computational time required for
estimating the CS-C-FV and CS-N models across different sample sizes
is presented in the Section~4.3 of the Supplementary Materials.

\section{Discussion} \label{sec:disc}

Building on previous research that modeled recurrent hypoglycemic
events using reflected Brownian motion to describe sequences hitting a
lower boundary \citep{xie2025recurrent}, we extended this framework to
a finite mixture FHT model to better capture the unobserved
heterogeneity in the patient population. This enhanced model
assumes the patient-level frailty in volatility to be
component-specific, allowing for distinct behavioral dynamics across
patient subgroups. To further investigate the relationships between
regression coefficients and frailty within the volatility context, we
examined several reduced models, each representing different patterns
of heterogeneity. Parameter estimation was carried out in a Bayesian
framework. The effectiveness and reliability of two Bayesian model
comparison criteria were evaluated through simulation studies, despite
the high computational cost.

Applying these models to hypoglycemic events from motivating data 
enabled us to evaluate the proposed methodology in a realistic
clinical trial setting and to explore the heterogeneity
among patients. 
An important implication of modeling latent heterogeneity is that
covariate effects may differ in both magnitude and direction across
latent subgroups. In such settings, a homogeneous model that targets
only a marginal effect may fail to detect statistically significant
associations, even when strong but oppositely signed effects are
present at the component level. This phenomenon is particularly
relevant in biomedical applications, where treatment responses often
vary substantially across unobserved risk groups. Motivated by that,
we further investigated this scenario through an additional simulation
study, which confirms that opposing component-specific effects can
indeed result in significance being masked under a non-mixture model.
Details of this simulation study are provided in the Supplementary
Materials Section~4.2.

Several model assumptions warrant discussion. The introduction of
an upper reflecting barrier imposes an upper limit on Brownian motion,
which effectively prevents glucose levels from reaching infinite and
eventually allows them to drop down. This assumption is reasonable for
several considerations. First, the phenomenon of glycosuria occurs
when glucose levels in diabetic patients become so elevated that they
surpass the kidneys' reabsorption capacity, leading to glucose being
excreted in the urine and a subsequent reduction in blood glucose
levels. Second, patients who consistently receive medication are able
to maintain their glucose levels within a desired range. Although
using a reflecting barrier to model the Brownian motion's reversal may
not precisely depict the underlying mechanism of the glucose movement,
it serves as a straightforward modeling approach that offers valuable
insights into risk factors influencing glucose dynamics.

Sensitivity analyses on the choice of the Brownian motion reset point
have been previously conducted by \citet{xie2025recurrent}, with the
current model fixed at $10$. These analyses showed that, although the
estimated intercepts for the upper reflecting barrier and the
volatility vary with the choice of starting point, the covariate
effect estimates remain stable.
In the present work, we further examined this issue by conducting an
additional sensitivity analysis in which the starting point $x_0$ was
treated as a normal random variable with mean~$10$ and
variance~$\theta_3$. The resulting covariate effect estimates remain
highly consistent with those from the main analysis. 
However, allowing the starting point to be random does not yield
additional inferential insight beyond the fixed-$x_0$ specification
and increases model complexity. Accordingly, we retain the CS-C-FV
model with a fixed starting point as the primary analysis, and report
the randomized-$x_0$ specification only as a sensitivity analysis.
Detailed results are provided in Section~3 of the Supplementary
Materials.

Enhancing the current model is possible from several perspectives, but
each entails substantial increases in complexity and computational
demands. Allowing frailty to arise from different distributions would
provide greater model flexibility, yet it would also introduce a large
number of latent variables. In a Bayesian framework, these latent
variables are treated as parameters, substantially increasing the
computational burden. 
The number of components in the mixture model, currently fixed at two,
could be evaluated using model comparison criteria; 
However, adding additional mixture components (e.g., $K = 3$) to the
proposed model, which already includes two subject-level random
parameters, one of which is component-specific, would substantially
increase the number of latent variables. This increase in the number
of parameters would further complicate the model and may lead to weak
identifiability, poor MCMC mixing in finite samples, and longer
computational time.
For further
exploration, Dirichlet Process Mixture Models \citep{race2021semi} may
provide a viable alternative by avoiding the need to fix the number of
mixture components a priori.
We have also explored linking
covariates to the mixture weights to classify patients into distinct
groups, with the goal of generating insights for personalized
medicine. Although this approach theoretically enhances the model’s
utility, in practice it complicated the model structure and made MCMC
convergence more difficult in our studies.

\appendix
\section{FHT Distribution of Reflected Brownian Motion} \label{sec:FHTdist}

For a no-drift Brownian motion~$X(t)$ with volatility~$\sigma$, let $\kappa$ be 
the upper reflection barrier~$\kappa > \nu$ and $\nu$ be the lower boundary. 
Suppose that $X(0) = x_0 \in [\nu, \kappa]$ is the starting point, 
the first time when $X(t)$ hits~$\nu$ is $\tau := \inf \{t > 0; X(t) = \nu \}$. 
The density and distribution function of $\tau$ are, 
respectively \citep{hu2012hitting},
\begin{align}
  \label{eq:dens}
  f(t|x_0, \nu, \kappa, \sigma) &= \sum_{n=1}^{\infty} c_n \lambda_n e^{-\lambda_n t}, & t > 0, \\
  \label{eq:dist}
  F(t|x_0, \nu, \kappa, \sigma) &= 1 - \sum_{n=1}^{\infty} c_n e^{-\lambda_n t}, & t > 0,
\end{align}
where for $n = 1, 2, \ldots$,
\begin{align*}
  \lambda_n= \frac{(2n-1)^2 \sigma^2 \pi^2 }{8(\kappa - \nu)^2},
  \mbox{ and }
  c_n       = \frac{(-1)^{n + 1}4 }{(2n-1)\pi}
                \cos(\frac{ (2n - 1)\pi(\kappa - x_0)}{2(\kappa - \nu)}).
\end{align*}
Note that $0 < \lambda_1 < \lambda_2 < \cdots$, $\lambda_n \to \infty$, 
and $\sum_{n=1}^{\infty} c_n = 1$. Both functions involve infinite
series, which present challenges to accurate evaluation of the
log-likelihood and efficient design of random number generation. 
To maintain numerical accuracy in implementing the
density and distribution functions, we do
not impose a fixed number of terms~$N$ on the summation.
 Instead, the infinite series is evaluated
using a tolerance-based stopping rule, where the summation is
terminated when the ratio of the current term to the cumulative sum is
less than or equal to a prespecified threshold, set 
to $10^{-10}$ in this study. A comprehensive sensitivity analysis
regarding the choice of tolerance in the evaluation of the density and
survival probability, as well as parameter estimation in data
application, is provided in Supplementary Materials Section~3.3.

An efficient rejection algorithm to sample
random number from the distribution has been proposed 
by \citet{xie2025recurrent}. The implementation has been released in a 
\proglang{R} package \pkg{reflbrown} \citep{reflbrown-package}.

\bibliographystyle{asa}
\bibliography{refs}

@article{aalen2001understanding,
  title =	 {Understanding the shape of the hazard rate: A
                  process point of view (with comments and a rejoinder
                  by the authors)},
  author =	 {Aalen, Odd O and Gjessing, H{\aa}kon K},
  journal =	 {Statistical Science},
  volume =	 16,
  number =	 1,
  pages =	 {1--22},
  year =	 2001,
  publisher =	 {Institute of Mathematical Statistics}
}

@article{buse2009durability,
  title =	 {{Dura}bility of Basal versus Lispro Mix 75/25
                  Insulin Efficacy ({Durable}) Trial 24-Week Results:
                  {S}afety and Efficacy of Insulin Lispro Mix 75/25
                  Versus Insulin Glargine Added to Oral
                  Antihyperglycemic Drugs in Patients with Type 2
                  Diabetes},
  author =	 {Buse, John B and Wolffenbuttel, Bruce HR and Herman,
                  William H and Shemonsky, Natalie K and Jiang,
                  Honghua H and Fahrbach, Jessie L and Scism-Bacon,
                  Jamie L and Martin, Sherry A},
  journal =	 {Diabetes Care},
  volume =	 32,
  number =	 6,
  pages =	 {1007--1013},
  year =	 2009,
  publisher =	 {Am Diabetes Assoc}
}

@article{charles2019analyze,
  title =	 {How to Analyze and Interpret Recurrent Events Data
                  in the Presence of a Terminal Event: {A}n
                  Application on Readmission after Colorectal Cancer
                  Surgery},
  author =	 {Charles-Nelson, Ana{\"\i}s and Katsahian, Sandrine
                  and Schramm, Catherine},
  journal =	 {Statistics in Medicine},
  volume =	 38,
  number =	 18,
  pages =	 {3476--3502},
  year =	 2019,
  publisher =	 {Wiley Online Library}
}

@book{cook2007statistical,
  title =	 {The Statistical Analysis of Recurrent Events},
  author =	 {Cook, Richard John and Lawless, Jerald F},
  year =	 2007,
  publisher =	 {New York: Springer}
}

@article{cryer2003hypoglycemia,
  title =	 {Hypoglycemia in Diabetes},
  author =	 {Cryer, Philip E and Davis, Stephen N and Shamoon,
                  Harry},
  journal =	 {Diabetes Care},
  volume =	 26,
  number =	 6,
  pages =	 {1902--1912},
  year =	 2003,
  publisher =	 {Am Diabetes Assoc}
}

@Article{devalpine2017programming,
  title =	 {Programming With Models: {W}riting Statistical
                  Algorithms for General Model Structures with
                  {NIMBLE}},
  journal =	 {Journal of Computational and Graphical Statistics},
  volume =	 26,
  issue =	 2,
  pages =	 {403-413},
  year =	 2017,
  author =	 {Perry {de Valpine} and Daniel Turek and Christopher
                  Paciorek and Cliff Anderson-Bergman and Duncan
                  {Temple Lang} and Ras Bodik},
  doi =		 {10.1080/10618600.2016.1172487},
}

@article{dey1997bayesian,
  title =	 {Bayesian Approach for Nonlinear Random Effects
                  Models},
  author =	 {Dey, Dipak K and Chen, Ming-Hui and Chang, Hong},
  journal =	 {Biometrics},
  volume =	 53,
  number =	 4,
  pages =	 {1239--1252},
  year =	 1997,
  publisher =	 {JSTOR}
}

@article{economou2015bayesian,
  title =	 {Bayesian Threshold Regression Model with Random
                  Effects for Recurrent Events},
  author =	 {Economou, P and Malefaki, S and Caroni, C},
  journal =	 {Methodology and Computing in Applied Probability},
  volume =	 17,
  number =	 4,
  pages =	 {871--898},
  year =	 2015,
  publisher =	 {Springer}
}

@article{fu2016hypoglycemic,
  title =	 {Hypoglycemic Events Analysis via Recurrent
                  Time-to-Event ({H}EART) Models},
  author =	 {Fu, Haoda and Luo, Junxiang and Qu, Yongming},
  journal =	 {Journal of Biopharmaceutical Statistics},
  volume =	 26,
  number =	 2,
  pages =	 {280--298},
  year =	 2016,
  publisher =	 {Taylor \& Francis}
}

@article{geisser1979predictive,
  title =	 {A Predictive Approach to Model Selection},
  author =	 {Geisser, Seymour and Eddy, William F},
  journal =	 {Journal of the American Statistical Association},
  volume =	 74,
  number =	 365,
  pages =	 {153--160},
  year =	 1979,
  publisher =	 {Taylor \& Francis}
}

@article{gelfand1994bayesian,
  title =	 {{B}ayesian Model Choice: {A}symptotics and Exact
                  Calculations},
  author =	 {Gelfand, Alan E and Dey, Dipak K},
  journal =	 {Journal of the Royal Statistical Society: Series B
                  (Methodological)},
  volume =	 56,
  number =	 3,
  pages =	 {501--514},
  year =	 1994,
  publisher =	 {Wiley Online Library}
}

@article{gelman1992inference,
  title={Inference from iterative simulation using multiple sequences},
  author={Gelman, Andrew and Rubin, Donald B},
  journal={Statistical science},
  volume={7},
  number={4},
  pages={457--472},
  year={1992},
  publisher={Institute of Mathematical Statistics}
}

@article{heidelberger1983simulation,
  title =	 {Simulation Run Length Control in the Presence of an
                  Initial Transient},
  author =	 {Heidelberger, Philip and Welch, Peter D},
  journal =	 {Operations Research},
  volume =	 31,
  number =	 6,
  pages =	 {1109--1144},
  year =	 1983,
  publisher =	 {INFORMS}
}

@book{hofert2018elements,
  title =	 {Elements of Copula Modeling with R},
  author =	 {Hofert, Marius and Kojadinovic, Ivan and
                  M{\"a}chler, Martin and Yan, Jun},
  year =	 2018,
  publisher =	 {Springer}
}

@article{hu2012hitting,
  title =	 {The Hitting Time Density for a Reflected Brownian
                  Motion},
  author =	 {Hu, Qin and Wang, Yongjin and Yang, Xuewei},
  journal =	 {Computational Economics},
  volume =	 40,
  number =	 1,
  pages =	 {1--18},
  year =	 2012,
  publisher =	 {Springer}
}

@article{jiang2018clinical,
  title =	 {Clinical Trajectories, Healthcare Resource Use, and
                  Costs of Diabetic Nephropathy among Patients with
                  Type 2 Diabetes: {A} Latent Class Analysis},
  author =	 {Jiang, Ruixuan and Law, Ernest and Zhou, Zhou and
                  Yang, Hongbo and Wu, Eric Q and Seifeldin, Raafat},
  journal =	 {Diabetes Therapy},
  volume =	 9,
  pages =	 {1021--1036},
  year =	 2018,
  publisher =	 {Springer}
}

@incollection{lee2003first,
  title =	 {First Hitting Time Models for Lifetime Data},
  series =	 {Handbook of Statistics},
  publisher =	 {Elsevier},
  volume =	 23,
  pages =	 {537--543},
  year =	 2003,
  booktitle =	 {Advances in Survival Analysis},
  editor =       {N. Balakrishnan and C.R. Rao},
  issn =	 {0169-7161},
  author =	 {Mei-Ling Ting Lee and George A Whitmore}
}

@article{lee2006threshold,
  title =	 {Threshold Regression for Survival Analysis:
                  {M}odeling Event Times by a Stochastic Process
                  Reaching a Boundary},
  author =	 {Lee, Mei-Ling Ting and Whitmore, George A},
  journal =	 {Statistical Science},
  volume =	 21,
  number =	 4,
  pages =	 {501--513},
  year =	 2006,
  publisher =	 {Institute of Mathematical Statistics}
}

@article{lee2008threshold,
  title =	 {A threshold regression mixture model for assessing
                  treatment efficacy in a multiple myeloma clinical
                  trial},
  author =	 {Lee, Mei-Ling Ting and Chang, Mark and Whitmore, GA},
  journal =	 {Journal of Biopharmaceutical Statistics},
  volume =	 18,
  number =	 6,
  pages =	 {1136--1149},
  year =	 2008,
  publisher =	 {Taylor \& Francis}
}

@article{lee2019survey,
  title =	 {A Survey of Threshold Regression for Time-to-Event
                  Analysis and Applications},
  author =	 {Lee, Mei-Ling Ting},
  journal =	 {Taiwanese Journal of Mathematics},
  volume =	 23,
  number =	 2,
  pages =	 {293--305},
  year =	 2019,
  publisher =	 {JSTOR}
}

@article{ma2021heterogeneous,
  title =	 {Heterogeneous Individual Risk Modelling of Recurrent
                  Events},
  author =	 {Ma, Huijuan and Peng, Limin and Huang, Chiung-Yu and
                  Fu, Haoda},
  journal =	 {Biometrika},
  volume =	 108,
  number =	 1,
  pages =	 {183--198},
  year =	 2021,
  publisher =	 {Oxford University Press}
}

@inproceedings{malefaki2015modelling,
  title =	 {Modelling times between events with a cured fraction
                  using a first hitting time regression model with
                  individual random effects},
  author =	 {Malefaki, S and Economou, P and Caroni, C},
  booktitle =	 {Theory and Practice of Risk Assessment: ICRA 5,
                  Tomar, Portugal, 2013},
  editor =       {Kitsos, C. and Oliveira, T. and Rigas, A. and Gulati, S.},
  pages =	 {45--65},
  year =	 2015,
  organization = {Springer}
}

@misc{nimbleusermanual,
  title =	 {{NIMBLE} User Manual},
  author =	 {Perry {de Valpine} and Christopher Paciorek and
                  Daniel Turek and Nick Michaud and Cliff
                  Anderson-Bergman and Fritz Obermeyer and Claudia
                  {Wehrhahn Cortes} and Abel Rodrìguez and Duncan
                  {Temple Lang} and Sally Paganin},
  howpublished = {https://r-nimble.org},
  year =	 2025,
  version =	 {1.4.0},
  note =	 {{R} package manual version 1.4.0},
  doi =		 {10.5281/zenodo.1211190},
}

@article{papastamoulis2016label,
  title={{label.switching}: {A}n R package for dealing with the label 
  switching problem in MCMC outputs},
  author={Papastamoulis, Panagiotis},
  journal={Journal of Statistical Software},
  volume={69},
  pages={1--24},
  year={2016}
}

@article{pennell2010bayesian,
  title =	 {{B}ayesian Random-Effects Threshold Regression with
                  Application to Survival Data with Nonproportional
                  Hazards},
  author =	 {Pennell, Michael L and Whitmore, GA and Ting Lee,
                  Mei-Ling},
  journal =	 {Biostatistics},
  volume =	 11,
  number =	 1,
  pages =	 {111--126},
  year =	 2010,
  publisher =	 {Oxford University Press}
}

@article{plummer2006coda,
  title =	 {{CODA}: {C}onvergence Diagnosis and Output Analysis
                  for {MCMC}},
  author =	 {Plummer, Martyn and Best, Nicky and Cowles, Kate and
                  Vines, Karen},
  journal =	 {R News},
  volume =	 6,
  number =	 1,
  pages =	 {7--11},
  year =	 2006
}

@article{qu2022identifying,
  title =	 {Identifying Patient Profiles for Developing Tailored
                  Diabetes Self-management Interventions: {A} Latent
                  Class Cluster Analysis},
  author =	 {Qu, Haiyan and Shewchuk, Richard M and Richman,
                  Joshua and Andreae, Lynn J and Safford, Monika M},
  journal =	 {Risk Management and Healthcare Policy},
  pages =	 {1055--1063},
  year =	 2022,
  publisher =	 {Taylor \& Francis}
}

@article{race2021semi,
  title =	 {Semi-parametric Survival Analysis via {D}irichlet
                  Process Mixtures of the {F}irst {H}itting {T}ime
                  Model},
  author =	 {Race, Jonathan A and Pennell, Michael L},
  journal =	 {Lifetime Data Analysis},
  volume =	 27,
  pages =	 {177--194},
  year =	 2021,
  publisher =	 {Springer}
}

@misc{reflbrown-package,
  title =	 {{reflbrown}: {R}ecurrent Event Modeling based on
                  First Hitting Time of {R}eflected Brownian Motion},
  author =	 {Yingfa Xie and Jun Yan},
  year =	 2024,
  howpublished = {https://github.com/YingfaX/reflbrown},
  note =	 {{R} package version 0.1.0}
}

@article{seaquist2013hypoglycemia,
  title =	 {Hypoglycemia and Diabetes: {A} Report of a Workgroup
                  of the {A}merican {D}iabetes {A}ssociation and the
                  {E}ndocrine {S}ociety},
  author =	 {Seaquist, Elizabeth R and Anderson, John and Childs,
                  Belinda and Cryer, Philip and Dagogo-Jack, Samuel
                  and Fish, Lisa and Heller, Simon R and Rodriguez,
                  Henry and Rosenzweig, James and Vigersky, Robert},
  journal =	 {Diabetes Care},
  volume =	 36,
  number =	 5,
  pages =	 {1384--1395},
  year =	 2013,
  publisher =	 {Am Diabetes Assoc}
}

@article{spiegelhalter2002bayesian,
  title =	 {{B}ayesian Measures of Model Complexity and Fit},
  author =	 {Spiegelhalter, David J and Best, Nicola G and
                  Carlin, Bradley P and Van Der Linde, Angelika},
  journal =	 {Journal of the Royal Statistical Society: Series B
                  (Statistical Methodology)},
  volume =	 64,
  number =	 4,
  pages =	 {583--639},
  year =	 2002,
  publisher =	 {Wiley Online Library}
}

@article{stephens2000dealing,
  title={Dealing with label switching in mixture models},
  author={Stephens, Matthew},
  journal={Journal of the Royal Statistical Society: 
        {S}eries B (Statistical Methodology)},
  volume={62},
  number={4},
  pages={795--809},
  year={2000},
  publisher={Wiley Online Library}
}

@article{whitmore1986first,
  title =	 {First-passage-time models for duration data:
                  Regression structures and competing risks},
  author =	 {Whitmore, George A},
  journal =	 {Journal of the Royal Statistical Society: Series D
                  (The Statistician)},
  volume =	 35,
  number =	 2,
  pages =	 {207--219},
  year =	 1986,
  publisher =	 {Wiley Online Library}
}

@article{whitmore2007modeling,
  title =	 {Modeling Low Birth Weights Using Threshold
                  Regression: {R}esults for {US} Birth Data},
  author =	 {Whitmore, G A and Su, Yi},
  journal =	 {Lifetime Data Analysis},
  volume =	 13,
  number =	 2,
  pages =	 {161--190},
  year =	 2007,
  publisher =	 {Springer Nature BV}
}

@article{xie2025recurrent,
  title =	 {Recurrent Events Modeling based on a Reflected
                  {B}rownian Motion with Application to Hypoglycemia},
  author =	 {Xie, Yingfa and Fu, Haoda and Huang, Yuan and
                  Pozdnyakov, Vladimir and Yan, Jun},
  journal =	 {Biostatistics},
  volume =	 26,
  number =	 1,
  pages =	 {kxae053},
  year =	 2025,
  publisher =	 {Oxford University Press}
}

\end{document}


\maketitle

\section{Additional Data Summary}

\begin{table}[tbp]
\centering
\caption{Descriptive statistics of the continuous covariates.
BMI, body mass index;
BP, blood pressure;
Min, minimum;
Max, maximum;
SD, standard deviation.}
\label{tab:datasumm}
\begin{tabular}{lccccc}
\toprule
Covariate & Min & Median & Max & Mean & SD \\
\midrule
\multicolumn{6}{l}{\textit{Glargine}} \\
Fasting glucose (mmol/l)          & 1.28  & 10.37 & 25.96 & 10.81 & 3.59 \\
Adiponectin (\textmu g/ml)        & 0.01  & 5.64  & 48.71 & 6.89  & 5.18 \\
Fasting insulin (mIU/L)           & 0.18  & 7.81  & 142.68& 10.47 & 10.90 \\
Height (cm)                       & 124.25& 166.14& 197.04& 166.19& 10.91 \\
BMI (kg/$\mbox{m}^2$)             & 15.88 & 31.40 & 53.96 & 31.74 & 6.16 \\
Diastolic BP (mmHg)               & 48.16 & 78.73 & 116.30& 78.23 & 9.40 \\
Systolic BP (mmHg)                & 89.17 & 130.23& 196.67& 131.41& 15.96 \\
Heart rate (beats per minute)     & 50.40 & 76.84 & 121.05& 76.87 & 9.81 \\
Duration diabetes (years)         & 0.04  & 8.27  & 36.82 & 9.53  & 5.98 \\
[1ex]
\multicolumn{6}{l}{\textit{LM 75/25}} \\
Fasting glucose (mmol/l)          & 0.23  & 10.55 & 24.47 & 10.78 & 3.83 \\
Adiponectin (\textmu g/ml)        & 0.04  & 5.45  & 49.01 & 7.06  & 5.82 \\
Fasting insulin (mIU/L)           & 0.18  & 8.16  & 89.85 & 10.29 & 8.49 \\
Height (cm)                       & 139.93& 166.86& 198.09& 166.77& 10.55 \\
BMI (kg/$\mbox{m}^2$)             & 16.11 & 31.16 & 62.62 & 31.66 & 6.23 \\
Diastolic BP (mmHg)               & 45.01 & 78.71 & 111.57& 78.28 & 9.53 \\
Systolic BP (mmHg)                & 47.26 & 129.77& 191.88& 131.71& 16.30 \\
Heart rate (beats per minute)     & 43.86 & 76.39 & 119.29& 76.70 & 9.82 \\
Duration diabetes (years)         & 0.03  & 8.94  & 39.48 & 10.00 & 6.34 \\
\bottomrule
\end{tabular}
\end{table}

Table~\ref{tab:datasumm} summarizes additional descriptive statistics
of the continuous covariates in the analysis data.

\section{MCMC Convergence} 

\begin{figure}
\centering
\includegraphics[width=\textwidth]{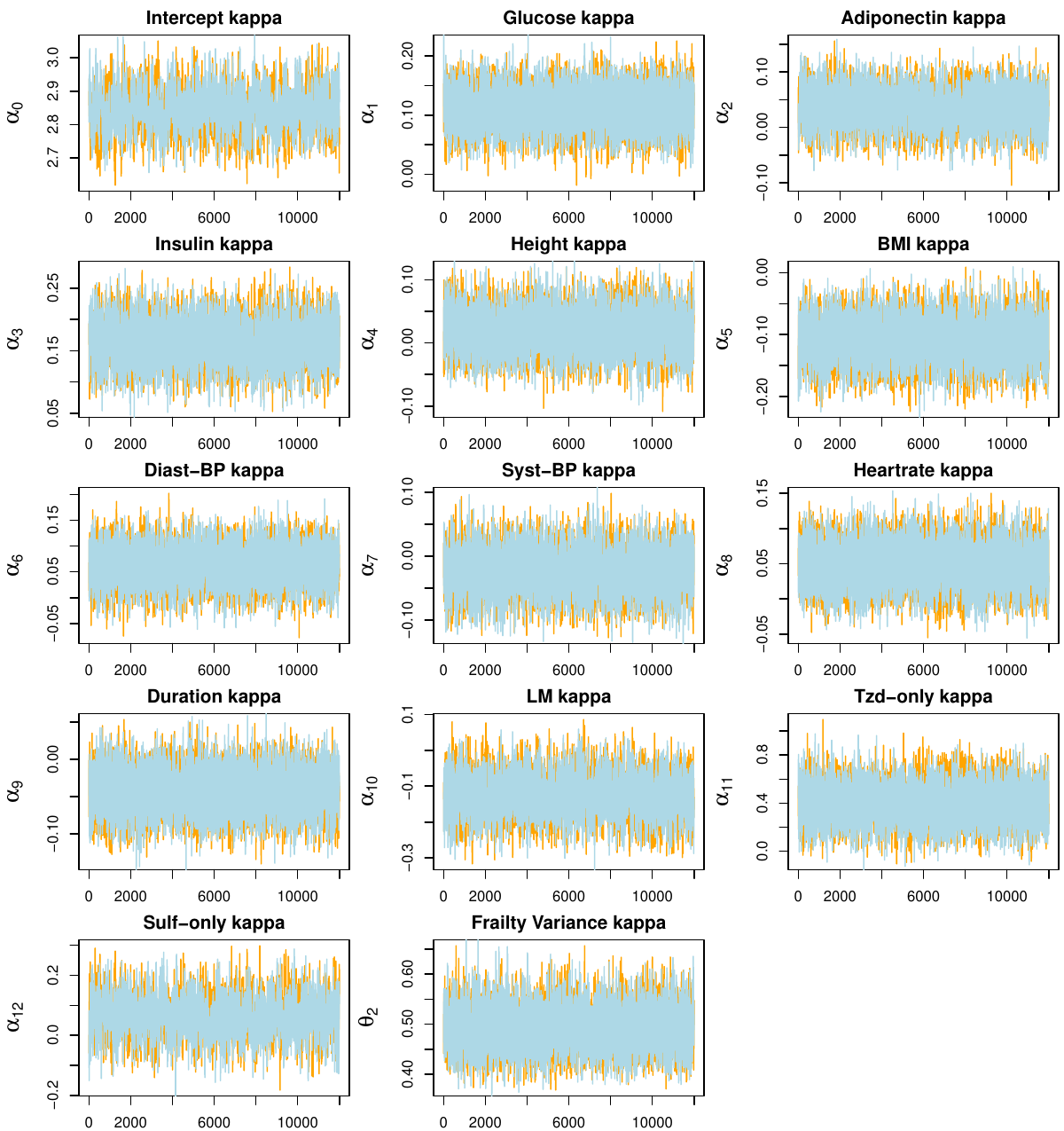}
\caption{Traceplots of the parameters in upper reflection barrier of
the CS-C-FV model}
\label{fig:tracep1}
\end{figure}

\begin{figure}
\centering
\includegraphics[width=\textwidth]{"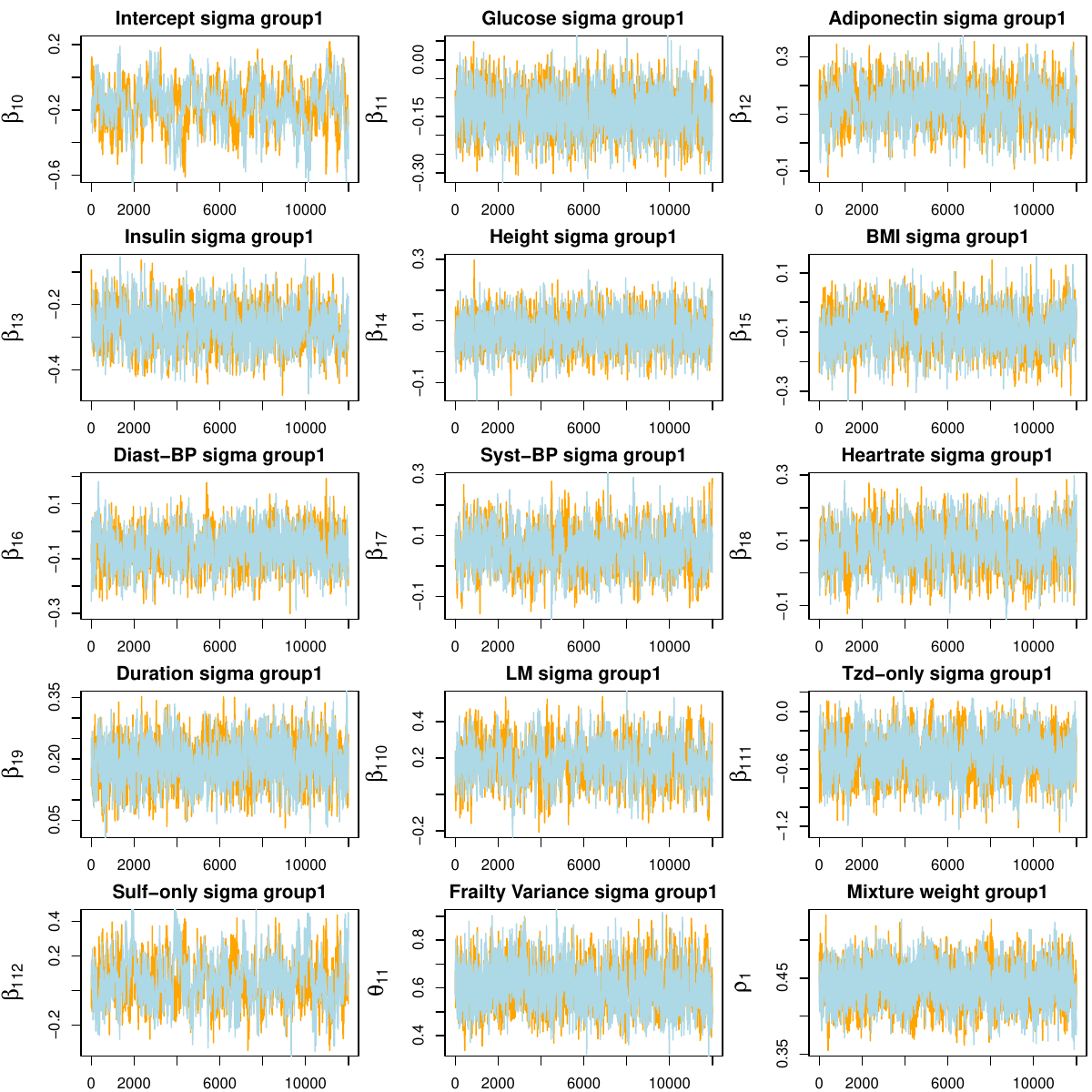"} 
\caption{Traceplots of the parameters in volatility of the first
component of the CS-C-FV model}
\label{fig:tracep2}
\end{figure}

\begin{figure}
\centering
\includegraphics[width=\textwidth]{"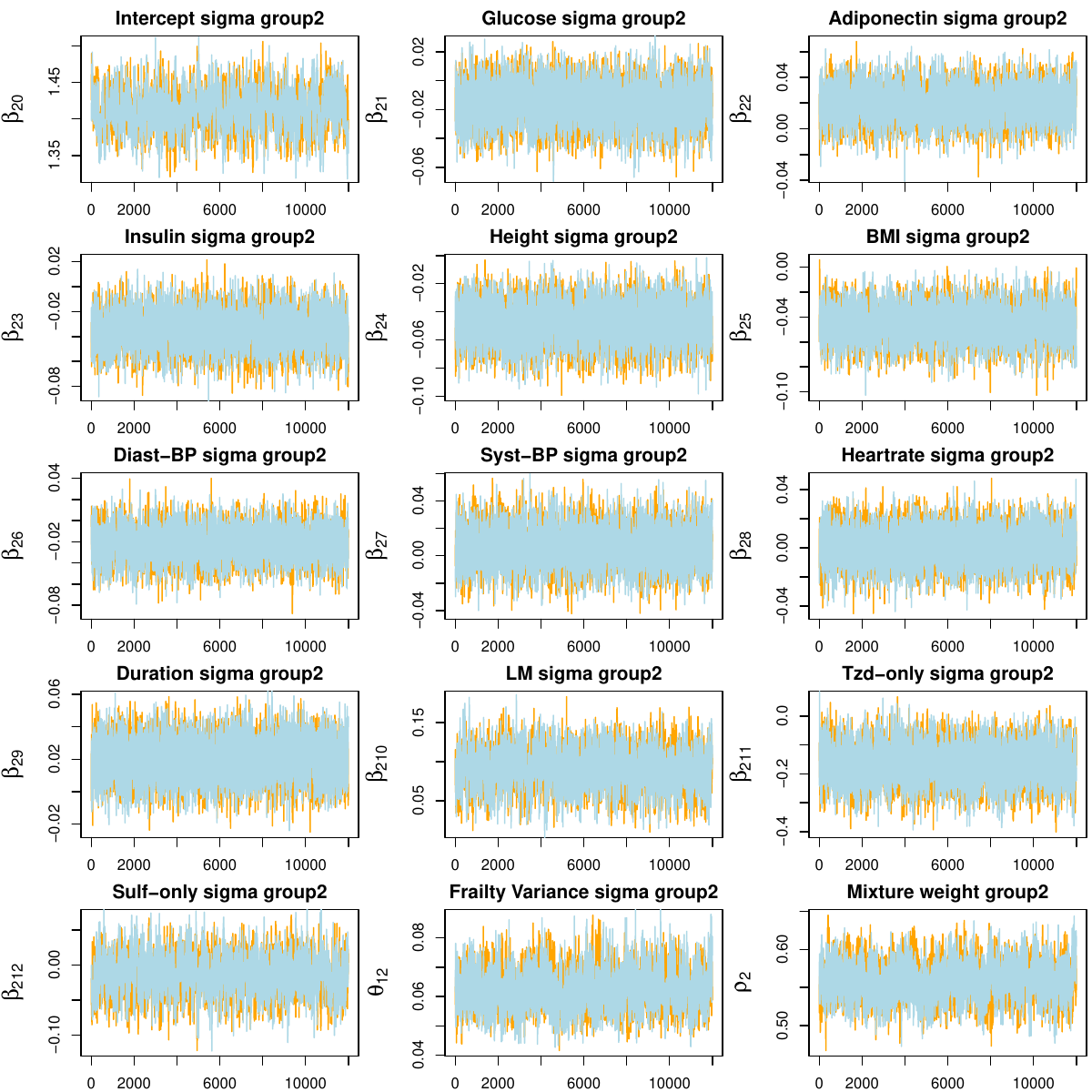"} 
\caption{Traceplots of the parameters in volatility of the second
component of the CS-C-FV model}
\label{fig:tracep3}
\end{figure}

The traceplots of the parameters of the
component-specific coefficient and frailty variance (CS-C-FV) model are
shown in Figure \ref{fig:tracep1} (corresponding to the parameters in
upper reflection barrier~$\kappa$), \ref{fig:tracep2} (corresponding
to the parameters in volatility~$\sigma$ of the first component), and
\ref{fig:tracep3} (corresponding to the parameters in
volatility~$\sigma$ of the second component). Two MCMC chains for each
parameter are run with different initial values. The traceplots show
that two MCMC chains mix well, indicating that the MCMCs have
converged.

\begin{table} [tbp]
\centering
\setlength{\tabcolsep}{5pt}
\caption{Convergence diagnostics for the CS-C-FV model. For each
parameter, the Gelman--Rubin potential scale reduction factor
($\hat{R}$), effective sample size (ESS), and Monte Carlo standard
error (MCSE) are reported for the posterior samples.
}
\label{tab:cvg_check_cscfv}
\begin{tabular}{llrllrllrl}
\toprule
				& \multicolumn{6}{c}{Volatility} 
				& \multicolumn{3}{c}{Upper reflection barrier} \\
				\cmidrule(lr){2-7}   \cmidrule(lr){8-10}
				& \multicolumn{3}{c}{Group 1} 
				& \multicolumn{3}{c}{Group 2} & \\
				\cmidrule(lr){2-4}   \cmidrule(lr){5-7}  
Covariates  	& $\hat{R}$  & ESS & MCSE
				& $\hat{R}$  & ESS & MCSE 
				& $\hat{R}$  & ESS & MCSE \\ 
\midrule
Intercept & 1.0064 & 193 & 0.0093 & 1.0010 & 742 & 0.0010 & 1.0192 & 967 & 0.0020 \\ 
Fasting glucose & 1.0019 & 1015 & 0.0016 & 1.0025 & 3411 & 0.0002 & 1.0039 & 4359 & 0.0005 \\ 
Adiponectin & 1.0012 & 671 & 0.0026 & 1.0000 & 2822 & 0.0002 & 1.0038 & 3142 & 0.0005 \\ 
Fasting insulin & 1.0003 & 755 & 0.0020 & 1.0001 & 2844 & 0.0003 & 1.0020 & 3286 & 0.0006 \\ 
Height & 1.0001 & 1098 & 0.0015 & 1.0015 & 3374 & 0.0002 & 1.0011 & 3963 & 0.0005 \\ 
BMI & 1.0048 & 809 & 0.0021 & 1.0001 & 3287 & 0.0002 & 1.0011 & 3322 & 0.0006 \\ 
Diastolic BP & 1.0028 & 810 & 0.0022 & 1.0019 & 2556 & 0.0003 & 1.0020 & 2802 & 0.0007 \\ 
Systolic BP & 1.0012 & 776 & 0.0023 & 1.0010 & 2564 & 0.0003 & 1.0008 & 3259 & 0.0005 \\ 
Heart rate & 1.0139 & 772 & 0.0022 & 1.0003 & 3364 & 0.0002 & 1.0002 & 4130 & 0.0004 \\ 
Duration diabetes & 1.0012 & 874 & 0.0016 & 1.0048 & 3567 & 0.0002 & 1.0008 & 4233 & 0.0004 \\ 
LM & 1.0084 & 547 & 0.0047 & 1.0006 & 1681 & 0.0006 & 1.0005 & 2025 & 0.0013 \\ 
tzd-only & 1.0004 & 620 & 0.0078 & 1.0002 & 2625 & 0.0012 & 1.0079 & 4060 & 0.0023 \\ 
sulf-only & 1.0053 & 303 & 0.0069 & 1.0000 & 1089 & 0.0008 & 1.0228 & 1157 & 0.0019 \\ 
Frailty std & 1.0262 & 890 & 0.0027 & 1.0083 & 1119 & 0.0002 & 1.0014 & 3463 & 0.0007 \\ 
Mixture Weight & 1.0073 & 776 & 0.0008 & 1.0073 & 776 & 0.0008 &  &  &  \\
\bottomrule
\end{tabular}
\end{table}

Additional, we compute the Gelman--Rubin
\citep{gelman1992inference} potential scale reduction factor using
multiple chains with \texttt{R} package \texttt{coda}
\citep{plummer2006coda}, and all parameters have values close to 1,
indicating good convergence. We report the effective sample
size (ESS) and the corresponding Monte Carlo standard error (MCSE) for
all model parameters. The ESS values are sufficiently large (relatively
small for some component-specific parameters), while the MCSE values
are small relative to the posterior standard deviations, suggesting
that Monte Carlo error has negligible impact on the posterior
summaries. These diagnostics collectively indicate satisfactory
convergence of the MCMC algorithm. The results are summarized in
Table~\ref{tab:cvg_check_cscfv}.

\section{Sensitivity Analysis}

\subsection{Sensitivity Analysis with a Random Starting Point}

We investigated an extension of the proposed framework by
allowing the starting point $x_0$ to be random. In the original
specification described in the main manuscript, the starting point
$x_0$ is fixed at $10$. To assess the sensitivity of the proposed
model to this assumption, we consider an alternative specification in
which $x_0$ is treated as a normal random variable by introducing a
subject-level random effect with a known mean~$10$ and
variance~$\theta_3$. This study allows us to examine whether
randomness in the starting point materially affects the estimation of
covariate effects compared with the fixed starting point setting
used in the main analysis.

\begin{figure}[htbp]
\centering
\includegraphics[width=\textwidth]{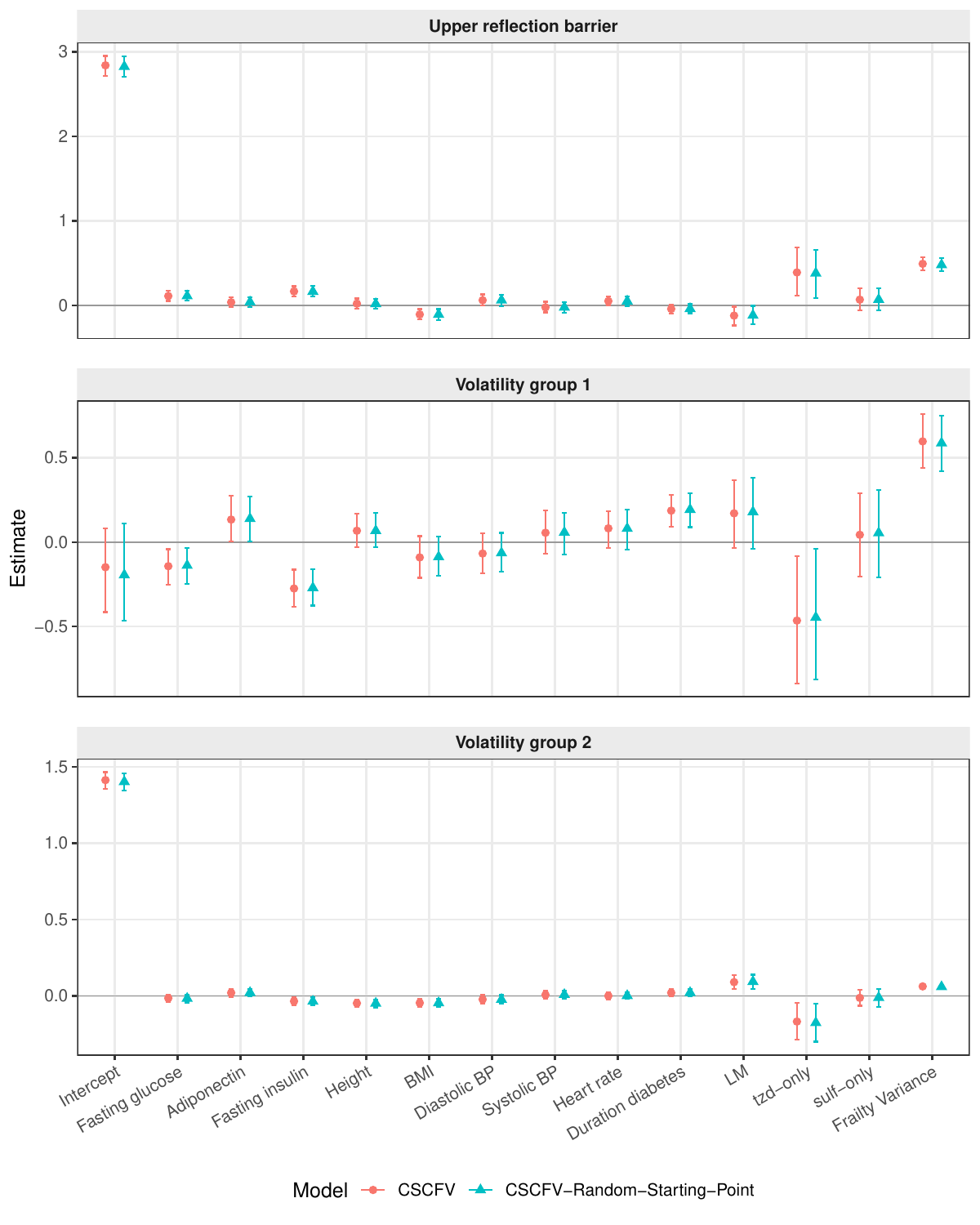}
\caption{Comparison of covariate effect estimates between the CS-C-FV
models with fixed and random starting points.}
\label{fig:est-CSCFV-x0-cpr}
\end{figure}

As shown in Figure~\ref{fig:est-CSCFV-x0-cpr}, the estimated covariate
effects from the CS-C-FV model with a random starting point are highly
consistent with those obtained under the fixed-$x_0$ specification.
The overall significance pattern of the covariate effects remains
unchanged, and other key parameter estimates are likewise very similar.
The estimated frailty standard deviation for the starting point is
0.53, corresponding to an estimated variance of $\theta_3 = 0.28$.
Taken together, these results indicate that the main findings of the
paper are robust to allowing subject-level variation in $x_0$, and
that this additional flexibility does not materially alter inference
for the current dataset.

Because introducing a random starting point does not provide additional
inferential insight beyond the fixed-$x_0$ specification while
increasing model complexity, we retain the fixed-$x_0$ CS-C-FV model as
the primary analysis in the manuscript. The randomized-$x_0$ extension
is therefore reported as a sensitivity analysis to demonstrate the
robustness of the proposed method to this modeling assumption, rather
than as the main modeling framework.

\subsection{Sensitivity Analysis with Initial Values for Mixture Weights}

\begin{table}[ht]
\centering
\caption{Sensitivity of class assignments to alternative starting
values for the mixture weights. Entries $i \rightarrow j$ represent
the number of subjects assigned to class $i$ in the primary run and
class $j$ in the alternative run. Agreement denotes the proportion of
subjects with identical class assignments across runs;
PS: primary setting; 
AS: alternative setting;
Est.: estimated.} 
\label{tab:SA_initial_mixweight}
\begin{tabular}{lccrrrrr}
\toprule
Setting & Starting Weights & Est. Weights &
1 $\rightarrow$ 1 & 
1 $\rightarrow$ 2 &
2 $\rightarrow$ 1 & 
2 $\rightarrow$ 2 & 
Agreement \\ 
\midrule
PS   & (0.4, 0.6) & (0.440, 0.560)  & - &   - &   - & - & - \\ 
AS 1 & (0.2, 0.8) & (0.440, 0.560)  & 845 & 2 & 3 & 1093 & 0.997 \\ 
AS 2 & (0.6, 0.4) & (0.443, 0.557)  & 846 & 1 & 7 & 1089 & 0.996 \\ 
AS 3 & (0.8, 0.2) & (0.438, 0.562)  & 844 & 3 & 1 & 1095 & 0.998 \\  
AS 4 & (0.5, 0.5) & (0.439, 0.561)  & 842 & 5 & 3 & 1093 & 0.996 \\  
\bottomrule
\end{tabular}
\end{table}

To assess the sensitivity of the inferred mixture component
assignments to the choice of starting values, we conducted a
sensitivity analysis using alternative initial values for the mixture
weights. In the primary MCMC run, the mixture weights~$\bm{\rho}$ were
initialized at $(0.4, 0.6)$. 
Four additional runs were performed with alternative starting values:
$(0.2, 0.8)$, $(0.6, 0.4)$, $(0.8, 0.2)$, and $(0.5, 0.5)$, 
representing both balanced and highly unbalanced initial
configurations. Across all runs, the posterior estimates of
$\bm{\rho}$ were nearly identical, indicating that the inference is
robust to the choice of starting values.

For each run, posterior samples of the latent class indicators were
obtained from the MCMC output. For each subject, we estimated the
posterior probability of belonging to each mixture component by
calculating the proportion of MCMC iterations in which the subject was
assigned to that component. Hard class assignments were then derived
by assigning each subject to the component with the larger posterior
probability (i.e., using 0.5 as the cut-off point).

We compared the resulting hard class assignments between the
alternative runs and the primary run.
Table~\ref{tab:SA_initial_mixweight} presents the cross classification
of assignments between the primary run and each alternative run. The
table entries represent the number of subjects assigned to class $i$
in the primary run and class $j$ in the alternative run. The agreement
rates were very high ($\ge 0.996$), indicating that the inferred class
assignments were stable across different starting values.

\subsection{Sensitivity Analysis for Truncation Level in Distribution Implementation}

\begin{figure}[htbp]
\centering
\includegraphics[width=\textwidth]{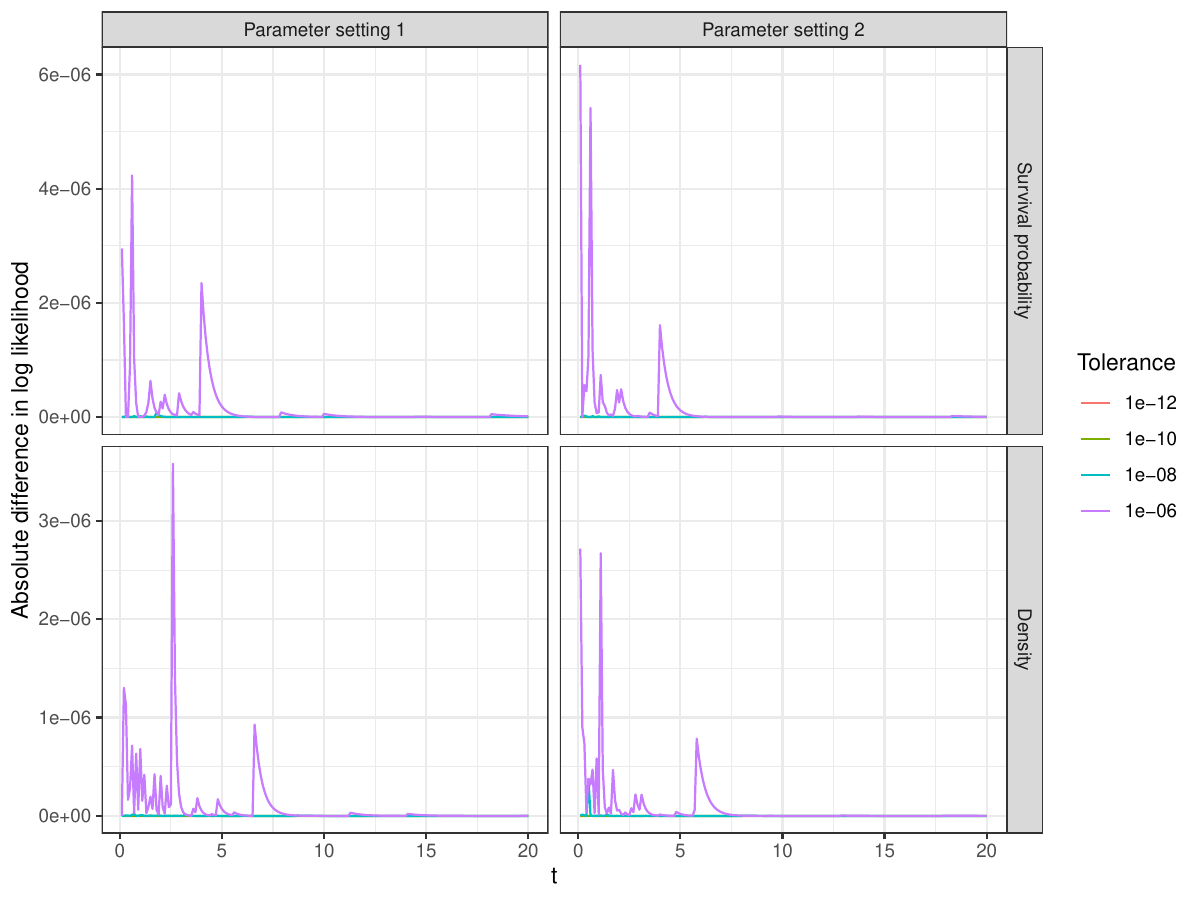}
\caption{Absolute differences in log-likelihood evaluation for the
FHT density and survival probability under different tolerance levels,
relative to the practically untruncated reference (tolerance $=0$).
Results are shown across representative parameter settings and gap
times. The parameter settings correspond to $(x_0, \nu, \kappa, \sigma)
= (10, 3.9, 20, 1)$ and $(10, 3.9, 30, 2)$.}
\label{fig:FHT_SA}
\end{figure}

\begin{figure}[htbp]
\centering
\includegraphics[width=\textwidth]{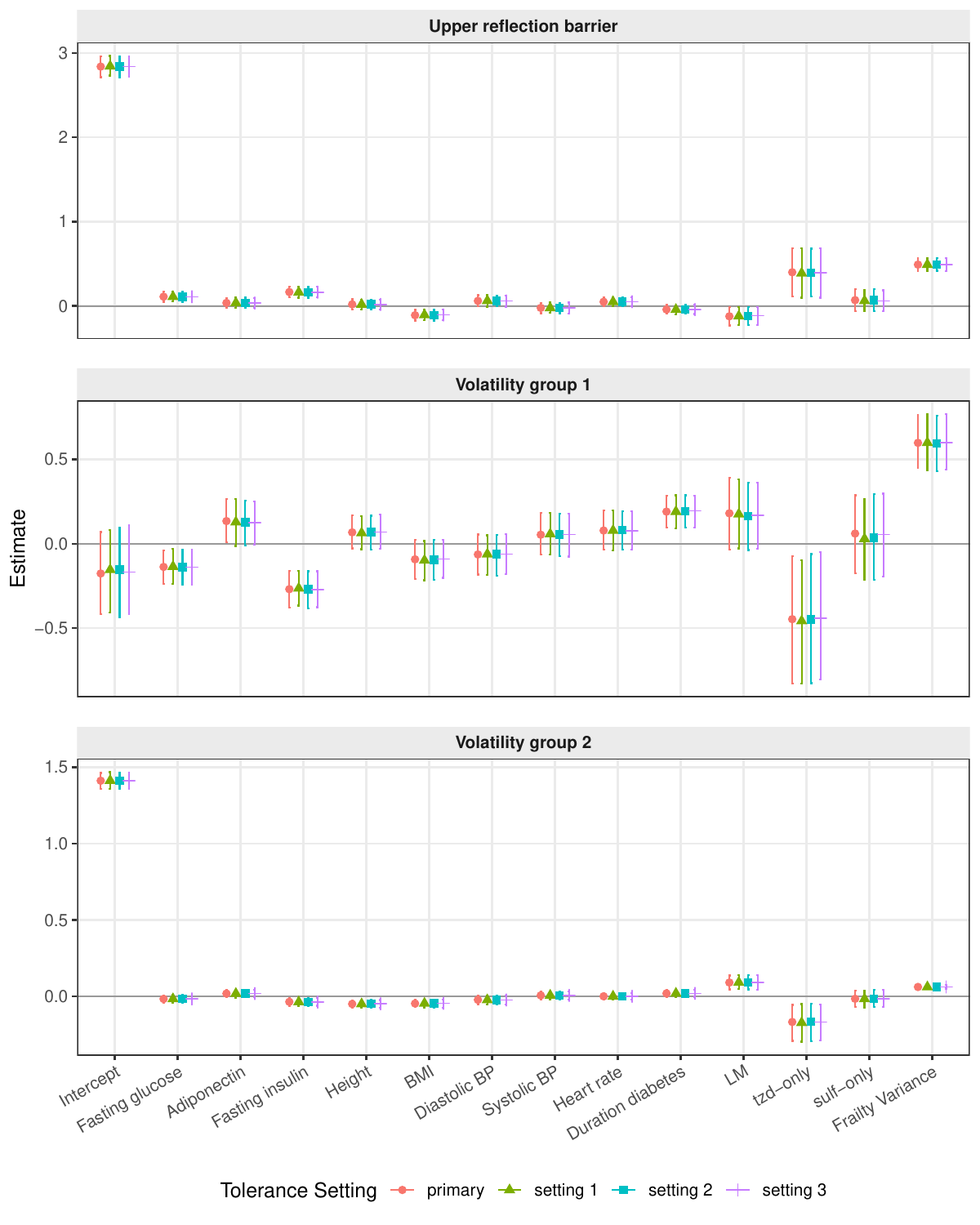}
\caption{Comparison of parameter estimates from the data analysis
under different tolerance levels used in the implementation of the FHT
density and distribution functions.}
\label{fig:est-CSCFV-tolerance-cpr}
\end{figure}

Additional sensitivity analyses for different tolerance levels in the
implementation of the reflected Brownian motion first hitting time
density and distribution functions \citep{hu2012hitting} were
conducted. The tolerance parameter controls the stopping rule for the
infinite series evaluation, where the summation is terminated when the
ratio of the current term to the cumulative sum is less than or equal
to a prespecified threshold.

We first examined the impact of the tolerance level on the numerical
evaluation of the FHT density and survival probability. Specifically,
we considered tolerance levels $0$, $10^{-6}$, $10^{-8}$, $10^{-10}$,
and $10^{-12}$, where tolerance $=0$ serves as a practically
untruncated reference. Figure~\ref{fig:FHT_SA} illustrates the
absolute differences in log-likelihood evaluation relative to this
reference across a range of representative parameter settings and gap
times. The differences are extremely small across all settings. Even
for the relatively coarse tolerance $10^{-6}$, the differences are at
the $10^{-6}$ scale or smaller, and are primarily observed when the
gap time is close to zero. As the gap time increases, the differences
rapidly diminish and become negligible. For tolerances $10^{-8}$ and
below, the results are virtually indistinguishable from the reference,
indicating that the stopping rule introduces minimal numerical error.

To assess the sensitivity of the estimation results to this numerical
criterion, we further repeated the data analysis under several
alternative tolerance levels. In addition to the primary setting
$(10^{-10})$, we considered three alternative values: $(10^{-6}),
(10^{-8})$, and $(10^{-12})$. These settings represent both looser and
stricter stopping criteria for the infinite series evaluation.

Figure~\ref{fig:est-CSCFV-tolerance-cpr} compares the parameter
estimates and their corresponding HPD intervals obtained under
these different tolerance levels. The results show that the estimates
remain highly consistent across all settings. Both the point estimates
and the confidence intervals exhibit negligible variation when the
tolerance parameter changes, indicating that the numerical
implementation of the infinite series is stable with respect to the
choice of truncation criterion.

\section{Additional Simulation Results}

\subsection{Smaller Sample Sizes}

\begin{table} [tbp]
\caption{Results of parameter estimation under correct specifications
	for three models with sample size $n \in \{200, 500\}$. 
	SD, posterior standard deviation; 
	ESD, empirical standard deviation; 
	CR, coverage rate of $95\%$ highest posterior density (HPD) 
	credible interval.}  
\label{tab:simuResSum}
\centering
\begin{tabular}{crrrrrrrrrrr}		
  \toprule
  &  &  & \multicolumn{4}{c}{$n = 200$} & \multicolumn{4}{c}{$n = 500$} \\
  \cmidrule(lr){4-7} \cmidrule(lr){8-11}
  Model & Para  & True & Bias & SD & ESD & CR & Bias & SD & ESD & CR \\ 
\midrule
CS-C-FV
& $\alpha_0$ & 2.90 & $-$0.025 & 0.091 & 0.089 & 0.94 & $-$0.015 & 0.056 & 0.056 & 0.94 \\ 
& $\alpha_1$ & 0.20 & $-$0.034 & 0.104 & 0.108 & 0.90 & $-$0.016 & 0.063 & 0.065 & 0.89 \\ 
& $\alpha_2$ & $-$0.10 & 0.019 & 0.087 & 0.091 & 0.90 & 0.005 & 0.054 & 0.053 & 0.93 \\ 
& $\beta_{10}$ & $-$0.20 & 0.302 & 0.430 & 0.549 & 0.90 & 0.031 & 0.167 & 0.169 & 0.94 \\  
& $\beta_{11}$ & $-$0.30 & 0.055 & 0.365 & 0.307 & 0.93 & $-$0.010 & 0.136 & 0.149 & 0.90 \\ 
& $\beta_{12}$ & $-$0.10 & 0.080 & 0.323 & 0.313 & 0.96 & 0.027 & 0.122 & 0.131 & 0.93 \\ 
& $\beta_{20}$ & 1.40 & 0.040 & 0.069 & 0.080 & 0.88 & 0.013 & 0.041 & 0.045 & 0.91 \\ 
& $\beta_{21}$ & $-$0.05 & 0.005 & 0.050 & 0.041 & 0.97 & $-$0.003 & 0.030 & 0.031 & 0.94 \\ 
& $\beta_{22}$ & $-$0.05 & 0.000 & 0.049 & 0.051 & 0.95 & $-$0.008 & 0.029 & 0.029 & 0.95 \\ 
& $\rho_1$ & 0.40 & 0.041 & 0.076 & 0.071 & 0.87 & 0.004 & 0.052 & 0.053 & 0.91 \\ 
& $\rho_2$ & 0.60 & $-$0.041 & 0.076 & 0.071 & 0.87 & $-$0.004 & 0.052 & 0.053 & 0.91 \\ 
& $\theta_{11}$ & 0.50 & 0.023 & 0.369 & 0.248 & 0.97 & 0.020 & 0.203 & 0.182 & 0.89 \\ 
& $\theta_{12}$ & 0.10 & $-$0.025 & 0.040 & 0.047 & 0.86 & $-$0.004 & 0.021 & 0.021 & 0.94 \\ 
& $\theta_2$ & 0.30 & $-$0.028 & 0.098 & 0.106 & 0.90 & $-$0.012 & 0.058 & 0.063 & 0.93 \\ 
\midrule
CS-I-FV
& $\alpha_0$ & 2.90 & $-$0.023 & 0.092 & 0.086 & 0.94 & $-$0.013 & 0.056 & 0.056 & 0.93 \\ 
& $\alpha_1$ & 0.20 & $-$0.025 & 0.104 & 0.109 & 0.92 & $-$0.013 & 0.063 & 0.066 & 0.91 \\ 
& $\alpha_2$ & $-$0.10 & 0.016 & 0.088 & 0.093 & 0.89 & 0.003 & 0.054 & 0.056 & 0.93 \\ 
& $\beta_{10}$ & $-$0.20 & 0.047 & 0.281 & 0.291 & 0.96 & 0.010 & 0.163 & 0.172 & 0.94 \\ 
& $\beta_{20}$ & 1.40 & 0.011 & 0.065 & 0.064 & 0.91 & 0.011 & 0.042 & 0.046 & 0.91 \\ 
& $\beta_{1}$ & $-$0.05 & 0.006 & 0.048 & 0.041 & 0.98 & $-$0.001 & 0.030 & 0.032 & 0.93 \\ 
& $\beta_{2}$ & $-$0.05 & 0.003 & 0.046 & 0.046 & 0.96 & $-$0.006 & 0.028 & 0.029 & 0.96 \\ 
& $\rho_1$ & 0.40 & 0.008 & 0.081 & 0.063 & 0.98 & 0.001 & 0.055 & 0.061 & 0.91 \\ 
& $\rho_2$ & 0.60 & $-$0.008 & 0.081 & 0.063 & 0.98 & $-$0.001 & 0.055 & 0.061 & 0.91 \\ 
& $\theta_{11}$ & 0.50 & 0.001 & 0.335 & 0.227 & 0.95 & 0.026 & 0.215 & 0.192 & 0.90 \\ 
& $\theta_{12}$ & 0.10 & $-$0.008 & 0.035 & 0.034 & 0.96 & $-$0.003 & 0.021 & 0.023 & 0.93 \\ 
& $\theta_2$ & 0.30 & $-$0.029 & 0.100 & 0.108 & 0.89 & $-$0.012 & 0.058 & 0.063 & 0.93 \\ 
\midrule
CS-I
& $\alpha_0$ & 2.90 & $-$0.024 & 0.093 & 0.091 & 0.95 & $-$0.012 & 0.057 & 0.059 & 0.94 \\ 
& $\alpha_1$ & 0.20 & $-$0.028 & 0.107 & 0.106 & 0.93 & $-$0.014 & 0.064 & 0.067 & 0.90 \\ 
& $\alpha_2$ & $-$0.10 & 0.014 & 0.090 & 0.096 & 0.92 & 0.004 & 0.055 & 0.056 & 0.94 \\ 
& $\beta_{10}$ & $-$0.20 & 0.011 & 0.119 & 0.115 & 0.97 & 0.004 & 0.073 & 0.074 & 0.94 \\ 
& $\beta_{20}$ & 1.40 & $-$0.007 & 0.048 & 0.046 & 0.97 & 0.001 & 0.030 & 0.034 & 0.92 \\ 
& $\beta_1$ & $-$0.05 & 0.006 & 0.046 & 0.036 & 0.99 & $-$0.002 & 0.029 & 0.030 & 0.90 \\ 
& $\beta_2$ & $-$0.05 & 0.001 & 0.044 & 0.046 & 0.94 & $-$0.006 & 0.027 & 0.028 & 0.95 \\ 
& $\rho_1$ & 0.40 & 0.004 & 0.044 & 0.035 & 0.95 & $-$0.000 & 0.028 & 0.029 & 0.95 \\ 
& $\rho_2$ & 0.60 & $-$0.004 & 0.044 & 0.035 & 0.95 & 0.000 & 0.028 & 0.029 & 0.95 \\ 
& $\theta_1$ & 0.10 & $-$0.006 & 0.025 & 0.025 & 0.95 & $-$0.002 & 0.015 & 0.013 & 0.97 \\ 
& $\theta_2$ & 0.30 & $-$0.032 & 0.101 & 0.114 & 0.89 & $-$0.013 & 0.059 & 0.064 & 0.93 \\ 
\bottomrule
\end{tabular}
\end{table}

Additional simulation studies are presented to examine performance at
smaller sample sizes. The main manuscript reports results for
$n \in {1000, 2000}$, whereas results for $n \in {200, 500}$ are
reported here.

Table~\ref{tab:simuResSum} shows that the estimation procedure remains
stable at these sample sizes. Parameter bias is small, estimated
standard deviations are close to the corresponding empirical standard
deviations, and coverage probabilities are close to the nominal level.

\subsection{Opposite Effect Directions in Mixture Components}

\begin{table}[ht]
\centering
\caption{True values in CS-C-FV model}
\label{tab:TrueformixVsnonmix}
\begin{tabular}{llrrrrr}
\toprule
Component & Parameter & True Value  \\ 
\midrule
Reflecting bound & $\alpha_0$ & 2.90  \\ 
& $\alpha_1$ & 0.20  \\ 
& $\alpha_2$ & $-$0.10  \\ 
[0.2em] 
Volatility Component 1
& $\beta_{10}$ & $-$0.20 \\ 
& $\beta_{11}$ & $-$0.40  \\ 
& $\beta_{12}$ & $-$0.05 \\
[0.2em] 
Volatility Component 2
& $\beta_{20}$ & 1.40  \\ 
& $\beta_{21}$ & 0.08  \\ 
& $\beta_{22}$ & $-$0.05 \\ 
[0.2em] 
Mixture Weight
& $\rho_1$ & 0.40  \\ 
& $\rho_2$ & 0.60  \\ 
[0.2em] 
Frailty Variance & $\theta_{11}$ & 0.50  \\ 
& $\theta_{12}$ & 0.10  \\ 
& $\theta_2$ & 0.30  \\ 
\bottomrule
\end{tabular}
\end{table}

\begin{figure}
\centering
\includegraphics[width=\textwidth]{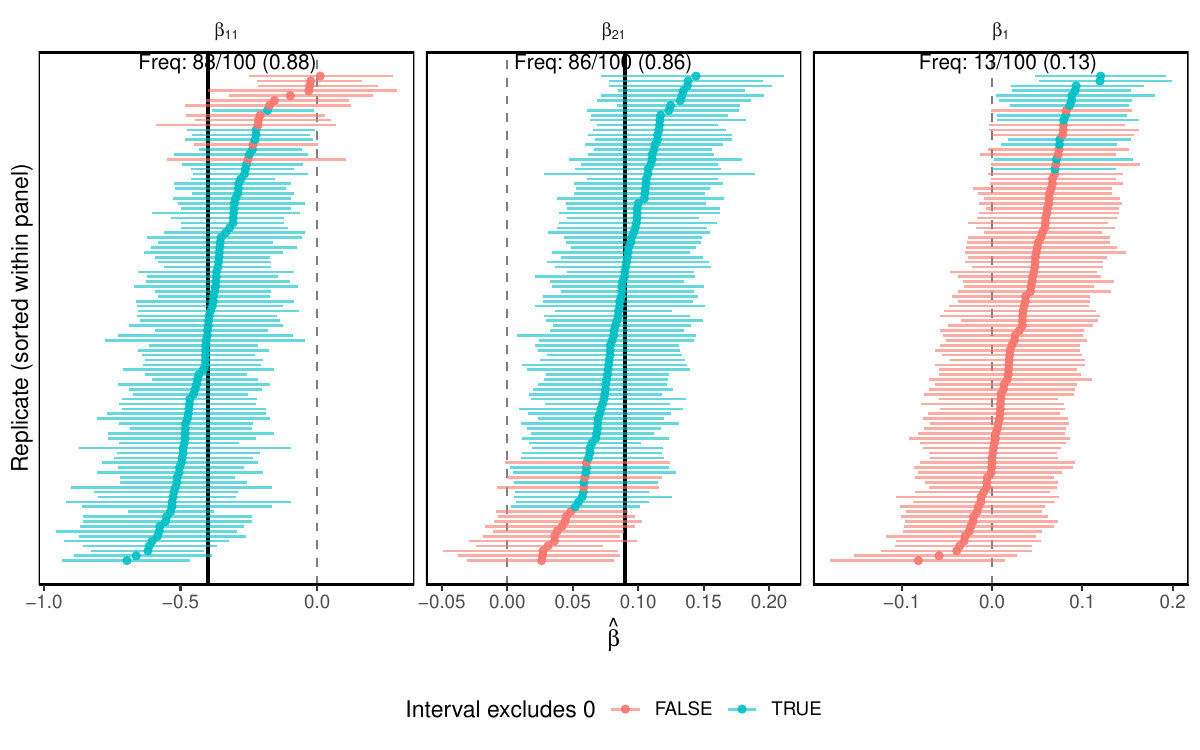}
\caption{Forest plots for estimation of $\beta$s}
\label{fig:mixVSnonmix}
\end{figure}

We conducted a simulation study to assess whether a non mixture model,
CS-N, can detect a treatment effect when the true treatment effects
have opposite signs across mixture components. Specifically, we set
the component specific treatment coefficients to $\beta_{11}=-0.4$ and
$\beta_{21}=0.08$, with mixing proportions $(\pi_1,\pi_2)=(0.4,0.6)$.
The true values of other model parameters are given in
Table~\ref{tab:TrueformixVsnonmix}. For each of 100 replicates, we
generated data from the proposed mixture model, CS-C-FV, fitted the
mixture model to obtain estimates and $95\%$ HPD credible intervals for
$\beta_{11}$ and $\beta_{21}$, and also fitted the non mixture model,
CS-N, to obtain an estimate and $95\%$ HPD credible intervals for the
treatment coefficient $\beta_1$. To visualize the results, we used
forest plots showing replicate level point estimates and HPD credible
intervals, with replicates sorted within each parameter panel.

Figure~\ref{fig:mixVSnonmix} summarizes the HPD credible intervals of
treatment effect estimates and corresponding $95\%$ HPD credible
intervals across
simulation replicates. When fitting the CS-C-FV, the component
specific treatment effects were detected with high empirical power.
Specifically, the treatment effect $\beta_{11}$ was statistically
significant in 88 out of 100 replicates (power = 0.88), and
$\beta_{21}$ was significant in 86 out of 100 replicates (power =
0.86), with HPD credible intervals concentrated around their
respective true values.

In contrast, the CS-N model exhibited substantially lower power for
detecting the treatment effect. The estimated coefficient $\beta_1$
was statistically significant in only 13 out of 100 replicates (power
= 0.13), with the majority of HPD credible intervals overlapping zero.
This pronounced difference reflects the cancellation of oppositely
signed component specific effects when aggregated into a single
marginal effect, highlighting the limitation of the non mixture model
in the presence of heterogeneous treatment responses.

\subsection{Model Estimation Runtime}

\begin{figure}
\centering
\includegraphics[width=\textwidth]{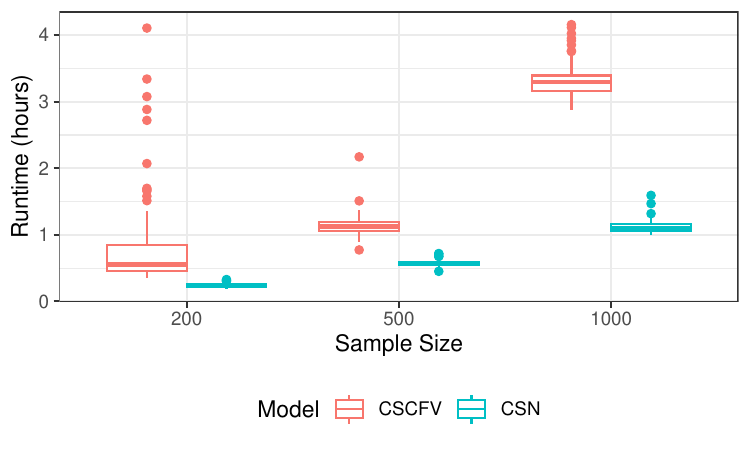}
\caption{Runtime for estimation of CS-C-FV and CS-N model}
\label{fig:runtimeCpr}
\end{figure}

Furthermore, we report the MCMC runtime (in hours) across 100
simulation replicates for each model at different sample sizes 
($n = 200, 500, 1000$), as shown in Figure~\ref{fig:runtimeCpr}. The
reported runtimes correspond solely to posterior sampling and exclude
the computation of model comparison criteria. Overall, the CS-C-FV
model requires longer computational time than the CS-N model, with the
difference becoming more pronounced as the sample size increases, as
expected given the additional model complexity of the CS-C-FV
specification.

Computing the model comparison criteria incurs substantially greater
computational cost, as it requires numerical integration via Monte
Carlo approximation at each posterior iteration. In our experiments,
the additional runtime associated with evaluating these criteria is
approximately 2.5 times the MCMC runtime.

\section{Computational Implementation}

Our implementation of the CS-C-FV model uses
the \texttt{R} package \texttt{NIMBLE}
\citep{devalpine2017programming}. After defining the model in NIMBLE
code, we adopt the most straightforward approach for running Markov
chain Monte Carlo (MCMC) sampling in NIMBLE by using the
\texttt{nimbleMCMC()} function.
This single function is used to define the underlying model and its
corresponding MCMC algorithm, compile both components, run the MCMC
procedure, and return posterior samples \citep{nimbleusermanual}.
NIMBLE automatically assigns sampler types based on model structure,
using Gibbs samplers when conjugacy is available and
Metropolis--Hastings or other generic samplers otherwise.

In our proposed model, the parameters of the FHT
distribution of the reflected Brownian motion do not have conjugate
priors, so NIMBLE assigns a random-walk sampler by default.
Specifically, this is an adaptive Metropolis--Hastings random-walk
sampler with a univariate normal proposal distribution \citep[see
Section 7.2.2.1 for details]{nimbleusermanual}. NIMBLE also allows
users to customize the MCMC configuration when needed.

Label switching is a well-known challenge in MCMC estimation for
Bayesian finite mixture models. We address the label switching issue
during the MCMC sampling process by imposing an identifiability
constraint on the component-specific parameters. Specifically, we
constrain the intercepts of the component-specific volatility
parameters~$\sigma$ such that $\beta_{10} \le \beta_{20}$. This
ordering constraint ensures that the two mixture components remain
identifiable throughout the MCMC sampling. This technique can be
implemented with NIMBLE 
\citep[see Section 5.2.7.3 for details]{nimbleusermanual}.

The implementation of the model with \texttt{NIMBLE} code has been
provided in a separately \texttt{R} file.

\bibliographystyle{asa}
\bibliography{refs}